\documentclass[12pt]{iopart}
\usepackage{iopams}  
\usepackage{graphicx}
\usepackage{cases}
\usepackage{url}
\usepackage{amssymb}

\usepackage{natbib}

\newcommand{\be}{\begin{equation}}
\newcommand{\ee}{\end{equation}}
\newcommand{\bea}{\begin{eqnarray}}
\newcommand{\eea}{\end{eqnarray}}
  
\newcommand{\pp}{\varphi}

\newcommand{\nn}{\nonumber}

\begin{document}

\title{Imaging a boson star at the Galactic center}

\author{F H Vincent$^{1,2}$, Z Meliani$^{3}$, P. Grandcl\'ement$^{3}$, E. Gourgoulhon$^{3}$, O Straub$^{3}$}
\address{$^{1}$ LESIA, Observatoire de Paris, PSL Research University, CNRS UMR 8109, Universit\'e Pierre et Marie Curie, Universit\'e Paris Diderot, 5 place Jules Janssen,
			92190 Meudon, France\\
	      $^{2}$ Nicolaus Copernicus Astronomical Center, ul. Bartycka 18, PL-00-716 Warszawa, Poland\\
        	      $^{3}$ LUTH, Observatoire de Paris, PSL Research University, CNRS UMR 8102, Universit«e Paris Diderot,
			5 place Jules Janssen, 92190 Meudon Cedex, France}
\ead{frederic.vincent@obspm.fr}

\begin{abstract}

Millimeter very long baseline interferometry will soon produce accurate images of the
closest surroundings of the supermassive compact object at the center of the Galaxy, Sgr~A*. These images
may reveal the existence of a central faint region, the so-called shadow, which is often interpreted as the
observable consequence of the event horizon of a black hole. In this paper, we compute
images of an accretion torus around Sgr~A* assuming this compact object is a boson star,
i.e. an alternative to black holes within general relativity, with no event horizon and no
hard surface. We show that very relativistic rotating boson stars produce images
extremely similar to Kerr black holes, showing in particular shadow-like
and photon-ring-like structures. This result highlights
the extreme difficulty of unambiguously telling the existence of an event horizon from strong-field images.

\end{abstract}

%\pacs{04.25.D-, 95.30.Sf}
%\maketitle

% INTRO %

\section{Introduction}

Kerr black holes are characterized by the existence of an event horizon, a surface
that separates the innermost region of spacetime from which no photons can reach a distant observer.
The image of the vicinity of a Kerr black hole surrounded by an optically thin accretion flow
is characterized by two specific features.
The central part of the image is dark because the black hole has by definition no emitting
surface and its event horizon captures photons traveling in the most central
parts of the spacetime.
This dark central area is known as the black hole 
\textit{shadow}~\citep{falcke00}\footnote{The term \textit{silhouette} is often used in place of
shadow. We keep here the original word, which describes properly the central fuzzy dark region
of strong-field images.}.
This shadow
is surrounded by a bright ring, the so-called \textit{photon ring}, made of photons winding
for one or many orbits in the very strong-field region extremely close to the black hole's event horizon.

The shape and angular size of the photon ring (or, equivalently, that of the shadow) contains very important information on the spacetime geometry
because it depends on the properties of the compact object. For a Kerr black hole, 
the shadow slightly changes with the observer's
inclination angle and with the black hole spin parameter~\citep{johannsen13}.
Many articles have investigated whether alternative compact objects exhibit
differences with respect to Kerr predictions~\citep{bambi09,johannsen10,amarilla10,amarilla13,vincent14,moffat15,cunha15}.

These two specific features of the Kerr black hole, the shadow and the photon ring, have
attracted considerable attention in the last few years because of the development of
millimeter Very Long Baseline Interferometry (VLBI). In particular, the 
\textit{Event Horizon Telescope}~\citep[EHT,][]{doeleman09}, which will become fully 
operational around 2020, will reach an angular resolution of $\approx 20~\mu$as.
This is less than the angular size of the shadow of the central black hole in our Galaxy, Sgr~A*,
which is $\approx 50~\mu$as, varying only slightly with the black hole spin. {We note that
the first EHT data were able to constrain the intrinsic angular size of the emitting region close
to Sgr~A* to only $37~\mu$as~\citep{doeleman08}}.
The shadow of the central black
hole of the galaxy M87 has an angular size of roughly half the size of Sgr~A* and is also a target of the EHT.
As a consequence, 
very near-future observations might allow constraining the Kerr metric parameters and
in particular the black hole spin from observing the size of the shadow of Sgr~A* and M87.
It might even be possible to constrain the actual theory of gravity in case the observed
shadow cannot be fitted by using the Kerr metric.

The capability of VLBI to demonstrate the existence of a shadow at Sgr~A*
was first advocated by~\citet{falcke00}. This reference put forward the fact
that detecting a shadow would be a proof of the existence of an event horizon.
Since then many articles have been investigating shadows and photon rings
in the perspective of the EHT (see the references given above).
These works are generally following one of three ways. They consider the
observable predictions of strong-field images:
\begin{itemize} 
\item either of a specific alternative theory of gravity, 
\item or of some specific alternative compact object within general relativity, 
\item or of some parameterization of the non-Kerrness of spacetime.
\end{itemize} 
The last way will
probably be the most efficient when analyzing an important set of data will be at stake.
However the two first ones are very important as well in order to determine how
specific to the Kerr metric the EHT observables are and in particular the existence
and angular size of the black hole shadow and photon ring.

This paper aims at developing the second way put forward above. We are interested
in determining the observable predictions of strong-field images of accretion flows
around \textit{boson stars}~\citep{feinblum68,kaup68,ruffini69}. These are alternative compact objects within the classical
theory of general relativity. Boson stars are particularly interesting as far as the future
EHT data are concerned because these objects have \textit{no event horizon and no emitting
hard surface}. They are thus perfect testbeds for examining whether shadows are
indeed a probe of the existence of an event horizon
and for determining the potential changes of a strong-field image caused by the absence of
such an event horizon, but still without any emitting hard surface~\citep[which is an important difference
with respect to an other well-studied alternative to black holes, the gravastar,][]{mazur04}.
This paper focuses on the particular case of the accretion flow surrounding Sgr~A*
as we have been investigating this environment in a recent work~\citep{vincent15}.

This work is one step in a series of paper aiming at examining the physical and astrophysical
properties of boson stars~\citep{grandclement14,meliani15}.

Section~\ref{sec:BStorus} presents boson stars and the accretion structure we consider.
Section~\ref{sec:results} gives our main results consisting in images and spectra of accretion tori
surrounding boson stars.
Section~\ref{sec:conclu} provides conclusions.

\section{Boson stars and accretion tori at Sgr~A*}
\label{sec:BStorus}

\subsection{Boson stars}
\label{sec:BSintro}

Boson stars are localized stable bundles of energy in the form of an assembly
of spin-$0$ bosons. The idea of a soliton-like distribution of energy kept
together by their own gravitational field dates back to the mid-$50$s with the
so-called geons (a particle-like solution of the coupled field equations of
general relativity and electromagnetism) developed by J. A. Wheeler~\citep{wheeler55}.
What is now called a boson star was developed by~\citet{feinblum68,kaup68,ruffini69}
who considered the Einstein-Klein-Gordon set of equations describing the
gravitational field created by an assembly of spin-$0$ bosons. Such boson stars
are macroscopic quantum objects subject to the Heisenberg uncertainty principle.
It is this principle that is at the basis of the fact that boson stars may not
undergo complete gravitational collapse to form a black hole.
%, provided they
%fulfill a stability condition (see the last paragraph of this Section). 
A lot of work has been devoted to these objects
and to their stability and we redirect to reviews containing the relevant
references~\citep{jetzer92,schunck03,liebling12}. 
%We note that these stability analyses have been developed in the context of
%spherically symmetric boson stars.

As of today, the only fundamental spin-$0$ boson is the Higgs boson detected
recently by the Large Hadron Collider. Should boson stars be made of Higgs
bosons, we would have to assume that the physical conditions inside these objects
make it possible for the Higgs decay processes and their reverse to reach an equilibrium,
in much the same way as for the $\beta$ decay in neutron stars.

A boson star is described by the Lagrangian
\be
L_{\mathrm{BS}} = L_g + L_\Phi
\ee
where $L_g$ is the Lagrangian of the gravitation field and $L_\Phi$ is the Lagrangian 
of a massive complex scalar field. Boson stars are objects described in the framework of classical
general relativity with minimal coupling of the scalar field. Accordingly
\be
L_g = L_{\mathrm{EH}} = \frac{1}{16 \pi G}R
\ee
is the standard Einstein-Hilbert Lagrangian, $R$ being the Ricci scalar.
The Lagrangian of the scalar field reads
\be
L_\Phi = -\frac{1}{2} \left(\nabla_\mu \Phi \nabla^\mu  \bar{\Phi} + \frac{m^2}{\hbar^2} |\Phi|^2\right)
\ee
where $\Phi$ is the complex scalar field. Boson stars are constructed by demanding it takes the form
\be
\Phi=\phi(r,\theta) \times \mathrm{exp}\left(i({\omega} t - {k} \pp)\right)
\ee
where $\phi$ being its modulus, $\omega$ its frequency
and the integer $k$ its azimuthal number. {Throughout this paper, unless otherwise stated,
we use quasi-isotropic coordinates $(t,r,\theta,\pp)$}. Note that although the scalar field
is time-dependent, the spacetime metric of boson stars is stationary. This is allowed by the
simple harmonic time dependence of the scalar field and by the fact that its energy-momentum
tensor only depends on the modulus of $\Phi$.

The parameter $m$ is the mass
of one individual boson which should not be confused with the total mass of the boson star.
In this framework, a boson-star spacetime is fully described by two parameters, the frequency
$\omega$ and the azimuthal number $k$. The boson mass $m$ is simply a scaling parameter,
in much the same way as the Kerr black hole mass. It can be shown that the pair $(\omega,k)$
should satisfy~\citep{grandclement14}
\bea
0 < \omega \le \frac{m}{\hbar}, \\ \nn
k \in \mathbb{N}.
\eea
The closer $\omega$ is to $m/\hbar$, the less relativistic (i.e. compact) is the boson star~\citep{grandclement14}. At the limit
of $\omega \rightarrow m/\hbar$, the scalar field vanishes. The boson star's 
angular momentum is directly
proportional to the azimuthal number~\citep{schunck98,grandclement14}
\be
J = k \hbar \mathcal{N}
\ee
where $\mathcal{N}$ is the total particle number of the boson star.
{Thus the angular momentum is simply proportional to $k$}. % increases with the star's rotation.
It is straightforward to compute a dimensionless spin parameter for a boson star in exactly the 
same way as for a Kerr black hole\footnote{The Kerr spin parameter is $a=J/M$ and has the dimension
of $M$. In this article we will consider the dimensionless quantity $a=J/M^2$ and call it spin for simplicity.}
\be
a = \frac{J}{M^2}
\ee
where $M$ is the total ADM (Arnowitt, Deser, Misner) mass of the boson star.
It is to be noted that contrarily to the Kerr black hole case, $a$ is not restricted to
be smaller than $1$~\citep{ryan97,grandclement14,meliani15}. {In the Kerr case,
the horizon is no longer defined for $a>1$ and the central singularity becomes naked.
As there is no event horizon nor a singularity for a boson star, nothing particular
occurs when $a>1$.}

We have not considered any self-interaction potential between the bosons, meaning that our
study is restricted to the so-called mini boson stars. We note that this restriction to mini boson
stars is important as far as astrophysical applications are concerned because the maximum
mass of a boson star is strongly 
%(precisely, in $\lambda^{1/2}$, where $\lambda$ is the
%interaction coupling constant) 
dependent on the existence or non-existence of interactions between
bosons~\citep{colpi86}. For a mini boson star with an azimuthal number of order a few, 
the total mass $M$ satisfies~\citep{grandclement14}
\be
M < M_{\mathrm{max}} = \alpha \frac{m_p^2}{m}
\ee
where $\alpha$ is a coefficient of order $1 - 10$ and $m_p$ is the Planck mass. For a Higgs boson ($m=125$~GeV), the
maximum mass is of order $10^{-21}\,M_{\odot}$, which is of course unable to account for
any black-hole-like astrophysical source. In order to get a total mass of $\approx 10^6\,M_{\odot}$
(of the order of the mass of Sgr~A*), the individual bosons should have a mass of $10^{-16}~$eV.
We note again that much higher masses (consistent with supermassive black holes) can be 
produced by taking self-interaction into account, without having to postulate the existence
of extremely light bosons~\citep{colpi86}. However, for the sake of simplicity, we do not
consider such an interaction in this paper. We thus assume the existence of very light spin-0
bosons in order to model Sgr~A* by a mini boson star. We also note that~\citet{amaro10}
has provided limits on $m$ based on dark matter models, which are not compatible with
the very small value assumed here. However, we do not try in this paper to model
self-consistently supermassive black holes and dark matter with the same scalar field.

It is not obvious to model black hole candidates of very different masses with the
same boson. Once the parameter $m$ is fixed, the total mass of the boson star
is restricted between $0$ and the maximum mass $M_\mathrm{max}$ introduced above. 
As a consequence it may seem that
if a boson light enough to model the most massive supermassive black holes
was existing, it would be possible to model with the same boson all black hole
candidates, whatever their mass (from stellar-mass to supermassive). However,
this is not obvious because the total mass of the boson star can become very
small with respect to $m_p^2/m$ only in the limit of $\omega \rightarrow m/\hbar$.
And as $\omega$ grows towards this limit, the distribution of the scalar field becomes
less and less compact and the spacetime becomes less and
less relativistic~\citep{grandclement14}. As a consequence it appears difficult to
model all black hole candidates with one common scalar field. It would probably
be even problematic to model all supermassive black hole candidates (with
masses from $\approx 10^6\,M_{\odot}$ to $\approx 10^{10}\,M_{\odot}$)
with one common boson given the large mass span. However, it is not very
likely that all black hole candidates in the Universe would be boson stars, it
is very possible that Kerr black holes would coexist with boson stars.
In this article, we will not
investigate this question any further and we only consider one object, Sgr~A*,
for which we chose the boson mass $m$.

Varying the action constructed from the Lagrangian $L_{\mathrm{BS}}$ with respect to the metric leads to
the usual Einstein field equations with the energy-momentum tensor of the scalar field.
Varying it with respect to the scalar field leads to Klein-Gordon equation. This set of equations
is solved using the KADATH library~\citep{grandclement10,grandclement14}. In this paper, we use the set
of metrics derived in~\citet{grandclement14}. We will consider only a few pairs of $(k,\omega)$
corresponding to few of the solutions illustrated in Figure~6 of~\citet{grandclement14}
and referenced in Table~\ref{tab:BS}. In particular,
we will not consider any boson-star spacetime containing an ergoregion as these solutions
are unstable~\citep{friedman78}. However, the timescale of the instability is not known and may be high enough
to allow such configurations to exist~\citep{grandclement14}. We will consider rotating boson stars with azimuthal
number $k=1$ and $k=4$, corresponding to the smallest and highest angular momentum of rotating
boson stars computed in~\citet{grandclement14}. We will consider three values of the frequency,
$\omega = 0.7, 0.8, 0.9\, m/\hbar$ spanning the spectrum from very relativistic ($\omega=0.7\,m/\hbar$)
to mildly relativistic ($\omega=0.9\,m/\hbar$) solutions. For both $k=1$ and $k=4$, an ergoregion starts
to develop for values of $\omega \lesssim 0.65\,m/\hbar$. We consider also non-rotating boson stars with
$k=0$, taking into account two values of the frequency $\omega=0.83, 0.9 \,m/\hbar$. For smaller values
of the frequency, two solutions exist for the same value of $\omega$~\citep{grandclement14} and we restrict ourselves
to the region of the parameter space with only one solution for one pair $(k,\omega)$. 
%the aim of this paper is not to provide a full overview of the parameter space of boson-star spacetimes,
%but rather to capture the main properties of few solutions spanning the parameter space.

We note that two of these spacetimes are secularly unstable.
Indeed, a curve $M(\omega)$ can be plotted for all values of $k$~\citep[see Fig.~6 of][]{grandclement14}.
At least for the smallest values of $k$, this curve shows a maximum for some value $\omega_\mathrm{max}(k)$.
A secular stability condition of the boson star is that $\omega>\omega_\mathrm{max}(k)$~\citep{friedman88}.
As $\omega_\mathrm{max}(k=4)$ is within the region of the $M(\omega)$ curve where an ergoregion
exists, the $k=4$ spacetimes considered here are all stable. However, $\omega_\mathrm{max}(k=0) \approx 0.86$
and $\omega_\mathrm{max}(k=1) \approx 0.77$, thus the $(k=0,\omega=0.83)$ and $(k=1,\omega=0.7)$
spacetimes are secularly unstable. We are still interested to investigate them in order to obtain a broad
range of boson-star images, with also very relativistic configurations (i.e. with small values of frequency).

\subsection{Accretion tori}
\label{sec:acctorus}

The dynamical evolution of normal (baryonic) matter accreted onto a boson star 
has not been much investigated in the past. \citet{torres02} dating
back to more than $10$ years ago seems to still contain the most developed discussion.
%\textbf{[Anyone knows other references??]}
It considers one of the most important questions, which is the possibility that accreting
matter, by concentrating to the center of the boson star, would create a black hole there
that could grow and ultimately encompass most of the scalar field distribution
below its horizon. 
Considering this problem, \citet{torres02} shows that if a supermassive boson star is present at
the Galactic center and accretes at the current rate during the age of the Universe,
it would still be $2$ orders of magnitude less massive than Sgr~A* (note that this computation
is using a very high value of the accretion rate in the innermost accretion flow - $10^{-6}\,M_{\odot}\,\mathrm{yr}^{-1}$- so it should be considered as an upper limit). 
This is an argument
in favor of the fact that should a supermassive boson star exist at Sgr~A*,
it could not have been turned into a black hole by accreting normal matter to its center. 
We also note that it is not straightforward that matter would be able to accumulate
at the center of the boson star: it should in particular be able to fight against a strong
angular momentum barrier.
Moreover, \citet{torres00} advocates the idea that stars accreted by a boson star at the
Galactic center would be fully disrupted by tidal effects and that the remaining
matter would end in unbound orbits, thus not accumulating at the center. However,
more work is needed in this area to get a clear picture of how accreted matter
would behave and how likely it is to form a black hole at the center of an accreting boson star. 
In this article, we consider a stationary toroidal accretion configuration
and we do not discuss its stability.

We model the accretion flow surrounding Sgr~A* by a constant-specific-angular-momentum,
circularly-orbiting, perfect-fluid, polytropic accretion torus.
We have already studied the properties of such accretion tori surrounding boson stars~\citep{meliani15}.
We combine here this work with
our recent model of an accretion torus surrounding a Kerr black hole at Sgr~A*~\citep{vincent15}.
Exactly the same model is used here, meaning that millimeter synchrotron radiation is emitted
by the optically thin accretion torus. We refer to~\citet{vincent15} for more details. The main difference
between the Kerr case and the boson star case, as far as accretion tori are concerned, is that there
does not always exist a self-crossing equi-pressure line (a cusp) in a boson-star spacetime~\citep{meliani15}.
In~\citet{vincent15} we assumed that the inner radius of the torus is located at the cusp (which always
exists for a Kerr black hole). In a boson star spacetime, we choose rather to let the inner radius be a free parameter.
As a consequence, an accretion torus surrounding a boson star is described by $9$ parameters (referenced in 
Tables~\ref{tab:kerrbf} and~\ref{tab:BS}):
the boson-star parameters $(\omega,k)$, the observer's inclination $i$, the constant angular momentum
$\ell = -u_\pp / u_t$ (where $\mathbf{u}$ is the fluid 4-velocity), the polytropic index $k_{\mathrm{p}}$, the inner
radius of the torus $r_{\mathrm{in}}$, the torus central temperature $T_c$ and number density $n_c$,
and the plasma $\beta$ parameter being the ratio of the gas to magnetic pressures.

\section{Images and spectra}
\label{sec:results}

In the whole paper, images and spectra of accretion tori surrounding
black holes and boson stars are computed using the open-source\footnote{Freely available at \url{http://gyoto.obspm.fr}} 
GYOTO code~\citep{vincent11}. 
Photons are traced backwards in time from a distant observer 
by integrating the geodesic equation
using a Runge-Kutta-Fehlberg adaptive-step
integrator at order 7/8 (meaning that the method is 8th order, with an
error estimation at 7th order). The integration is performed in the
Kerr metric (Section~\ref{sec:kerr}) or in the numerical spacetime of
a boson star computed by the KADATH library (Section~\ref{sec:BS}).
The equation of radiative transfer is integrated inside the optically thin torus
to determine the value of specific intensity reaching the observer in each direction
on sky (i.e. in each pixel of the observer's screen).

\subsection{Accretion tori around a Kerr black hole}
\label{sec:kerr}

\subsubsection{Reference Kerr image}
\label{sec:refkerr}

This section is meant to define a "reference" image of an accretion torus
surrounding a Kerr black hole, which will be used to interpret the subsequent
boson-star images. This setup is not the result of a proper fit, it is only a
set of parameters which allows to reasonably account for the observable
constraints that we currently have on the angular size of the emitting region
at $\lambda = 1.3$~mm and on the millimeter spectrum of Sgr~A*.

\begin{figure}
	\centering
	\includegraphics[height=6cm,width=13cm]{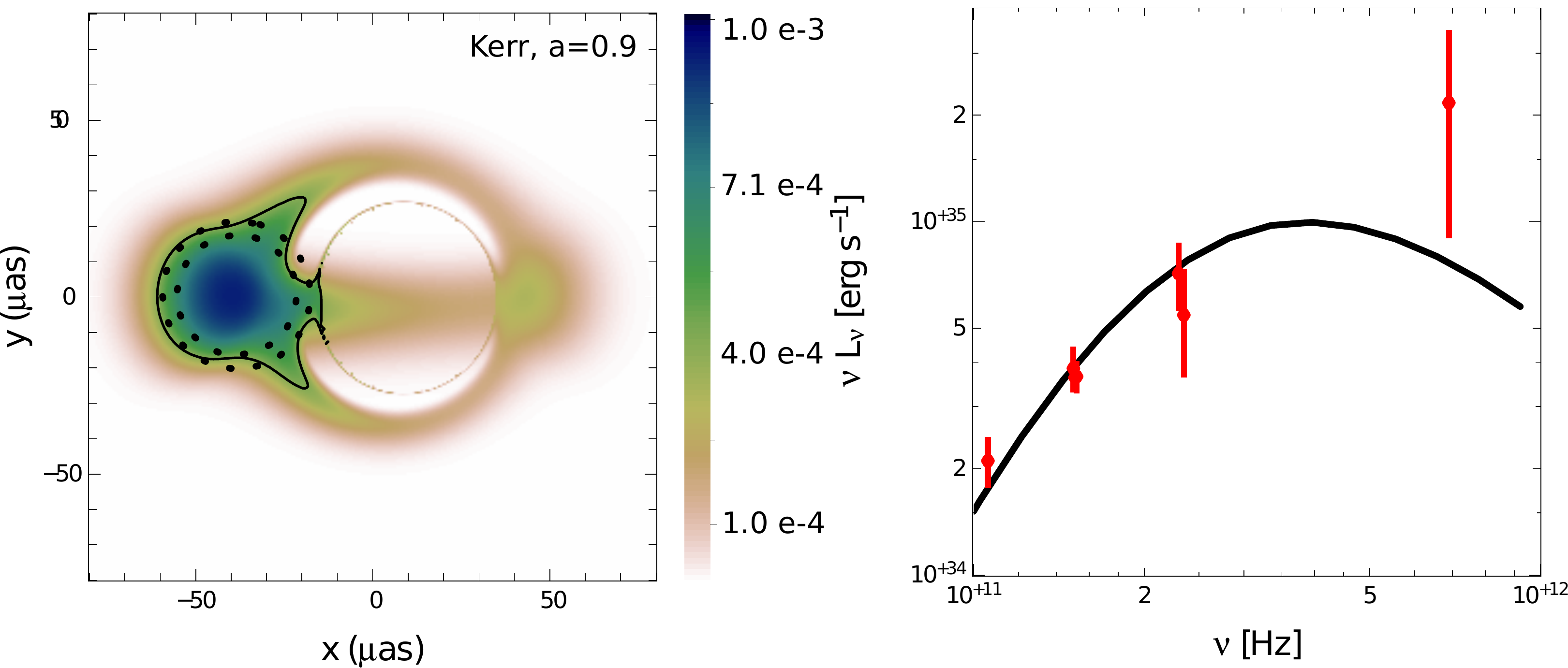} % \hspace{0.5cm}
	\includegraphics[height=6cm,width=6cm]{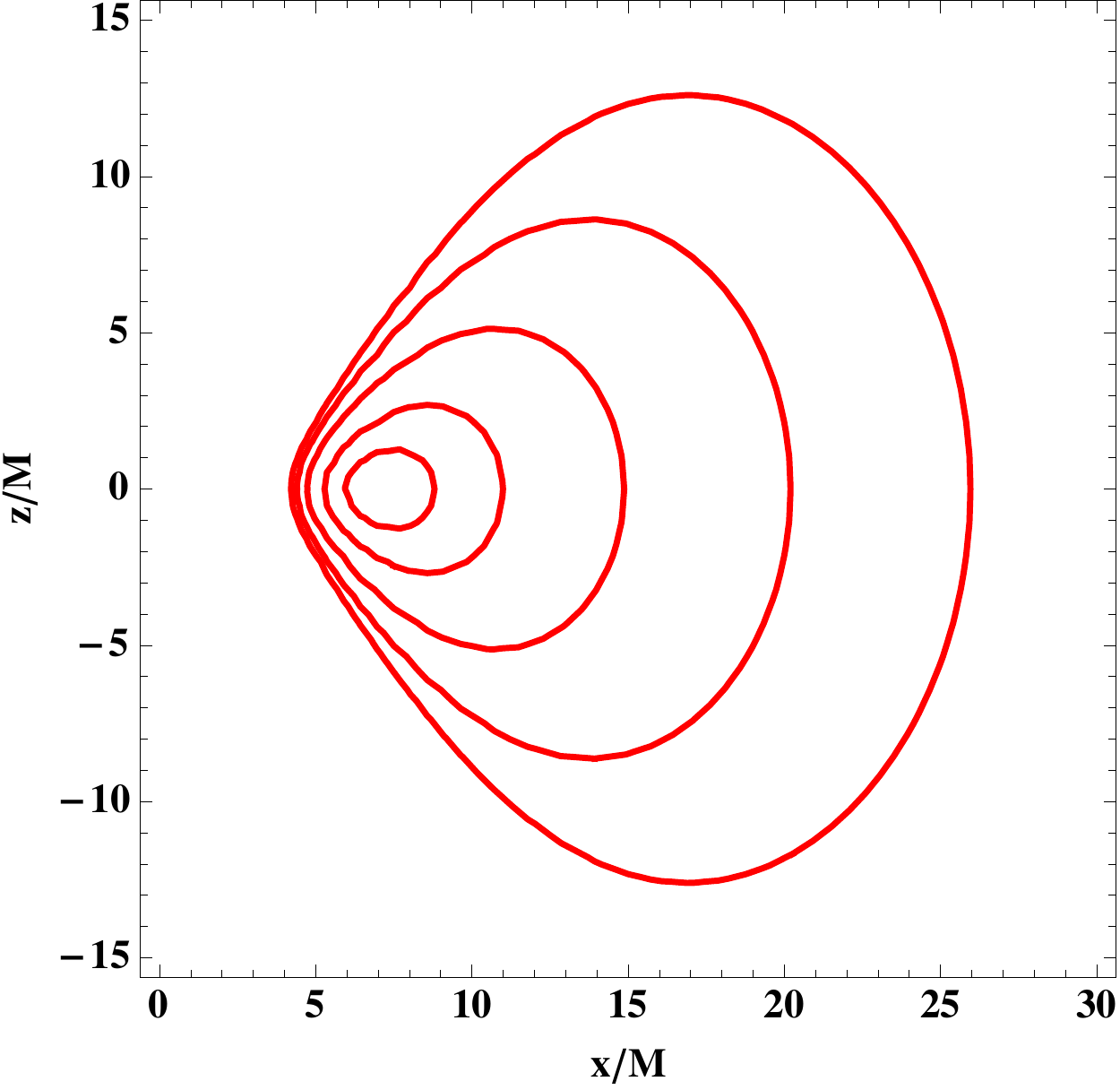}
	\caption{\textit{The reference Kerr case}. \textbf{Upper left:} image at $\lambda=1.3$~mm of an accretion torus surrounding a Kerr black hole with the parameters
	given in Table~\ref{tab:kerrbf}. The color bar indicates the cgs value of specific intensity. The dotted circles show the
	$1\sigma$ confidence limit for the intrinsic (no scattering included) angular size of the emitting zone~\citep{doeleman08}. 
	The solid black contour encompasses the region of the accretion flow emitting $50 \%$ of the total flux.
	\textbf{Upper right:} millimeter spectrum of the accretion torus, with red data points from~\citep{falcke98,marrone06}.
	\textbf{Lower panel:} equi-pressure contours of the accretion torus in the $(x,z)$ plane, $z$ being along the rotation axis.}
	\label{fig:kerrref}
\end{figure}

%Because of the discrete nature of their angular momentum,
%boson star rotates fast, as soon as they rotate. 
{Typical spin parameters of boson stars are close to $1$.}
The slowest-rotating {(k=1)} boson stars
that we analyze here have spin parameters of 
order $a \approx 0.9$~\citep{meliani15}. As a consequence, we consider a Kerr
spacetime with spin parameter $a=0.9$. Table~\ref{tab:kerrbf} shows the list of parameters
which allows to get a reasonable fit in the Kerr spacetime. It leads to the $1.3~$mm
strong-field image and to the millimeter spectrum shown in Fig.~\ref{fig:kerrref}. This Figure
also shows the equi-pressure contours of the reference Kerr torus. {We note in particular
that the radial extent of the torus is of order $20\,M$}.
\begin{table}[h!]
\centering                          % used for centering table
\begin{tabular}{c c c c c c c c}
\\
\hline \hline
\\
$a$	  & $\mathbf{i}$ & $\boldsymbol{\ell}$  & $r_{\mathrm{in}}$ & $\mathbf{n_c}$ & $\mathbf{T_c}$ & $\mathbf{k_{\mathrm{p}}}$ & $\boldsymbol{\beta}$
\\
\hline \hline
\\
$0.9$    &  ${85^{\circ}}$ & ${3.2}$~{M} & $4.2$~M & $6.3\times10^6$~$\mathrm{cm}^{-3}$ & ${5.3\times10^{10}}$~K & 5/3 & 10\\
\hline        \hline                                   %inserts single line
\end{tabular}
\caption{Parameters (introduced in Sect.~\ref{sec:acctorus}) used to fit the spectral and imaging constraints in the Kerr spacetime.
{We remind that $a$ is the dimensionless spin parameter, $i$ is the observer's inclination angle, 
$\ell$ is the fluid angular momentum $-u_\pp/u_t$, $r_{\mathrm{in}}$
is the torus inner radius, $n_c$ and $T_c$ are the torus central density and temperature, $k_\mathrm{p}$ is the polytropic index
and $\beta$ is the ratio of gas to magnetic pressures.}
Parameters in bold font will be kept fix in the whole paper. Only the spin parameter and the inner torus radius will vary.}
\label{tab:kerrbf}
\end{table}
We are interested in this paper in the modification of the strong-field image when the spacetime
is changed. As a consequence, we will keep fixed to their Kerr values given in Table~\ref{tab:kerrbf} 
all the astrophysical parameters $(\ell,n_c,T_c,k_{\mathrm{p}},\beta)$ together
with the inclination parameter $i$. The inner radius must be varied because different boson-star
spacetimes lead to tori with different radial extension for the same value of $r_{\mathrm{in}}$,
so that keeping the same value of $r_{\mathrm{in}}$ would have lead to very different looking images.
{We note that the total mass of the accretion torus is by many orders of magnitude smaller
than the mass of Sgr~A* which justifies the fact that we do not consider its contribution to the metric.}

At $1.3~$mm, interstellar scattering is still important~\citep{bower06} and will degrade the image with respect
to what is shown in Figure~\ref{fig:kerrref}, essentially convolving it with a Gaussian of FWHM$\approx 20 \,\mu$as.
In this analysis, we assume that this effect can be fully corrected (see~\citep{fish14} for a recent discussion).

\begin{figure}
	\centering
	\includegraphics[height=5cm,width=10cm]{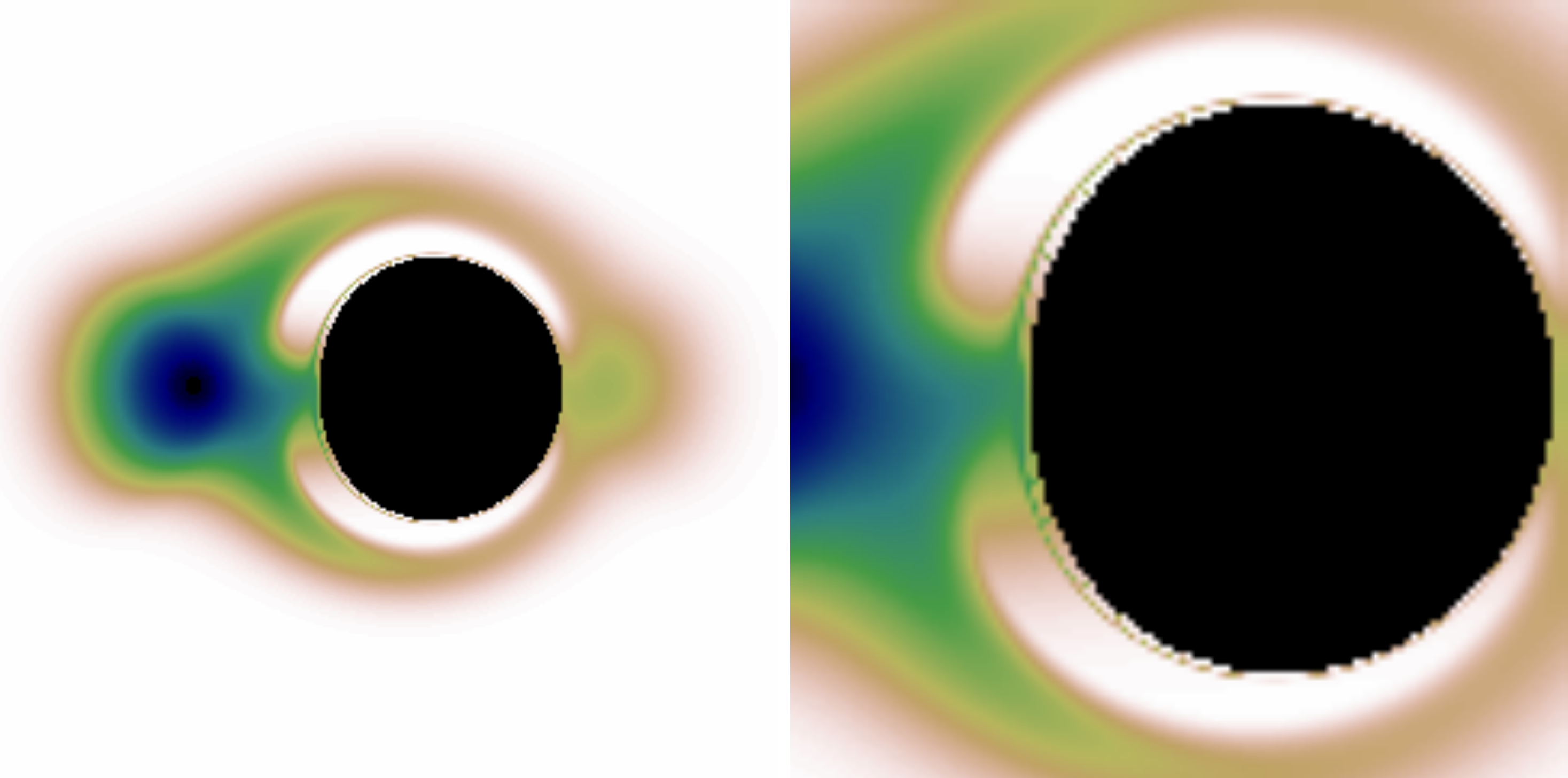} \hspace{0.5cm}
	\caption{\textit{Shadow and photon ring.} \textbf{Left:} same image as the left panel of Figure~\ref{fig:kerrref}, but the
	directions on the observer's sky that asymptotically approach the event horizon when ray tracing
	backwards in time are marked in black color. The black area at the center of the image is
	the black hole shadow. Its exterior limit nearly coincides with the photon ring.
	\textbf{Right:} zoom on the central region.}
	\label{fig:shadow}
\end{figure}

Figure~\ref{fig:kerrref} illustrates the notions of shadow and photon ring introduced earlier.
The photon ring is the bright nearly circular ring of light at the center of the image. It is nearly
exactly the outer limit of the black hole shadow, i.e. the locus of the directions on the observer's
sky that asymptotically approach the event horizon when ray tracing backwards in time\footnote{We note that geodesics
ray traced backwards in time should never cross the event horizon as this would correspond to geodesics
escaping the horizon, which is of course impossible. There is a stop condition in our ray-tracing code to prevent
infinite integration when a photon approaches "too close" to the horizon.}. 
Figure~\ref{fig:shadow} gives a precise illustration of
the location of the shadow. Comparing Figures~\ref{fig:kerrref} and~\ref{fig:shadow} shows
that the locus of the shadow is still illuminated in some parts because some radiation emitted
by the accretion torus in between the compact object and the observer will fall inside the shadow when projected on sky.
However, a strong gradient of specific intensity should be visible at least in some parts of the photon ring
\citep[particularly away from the equatorial plane and from the part of the image boosted by the relativistic
beaming effect, see][]{psaltis14}. In our model, matter is not emitting down to the event horizon: the inner
edge of the torus ($r_{\mathrm{in}} = 4.2\,M$) is the closest region where radiation is emitted. This is a condition for getting such a clear
photon ring as illustrated in Figure~\ref{fig:kerrref}. However, even in case matter is emitting all the way down to the event horizon,
there is still a sharp transition between the shadow and the outer region, as illustrated e.g. in Figure~1
of~\citet{falcke00}. As a consequence, it is really the strong gradient at the limit of the shadow
which is the observable of interest, whatever the astrophysical model. 
Demonstrating the existence and measuring the angular size 
of this shadow (and of the surrounding photon ring if visible)
is the main target of the EHT as far as strong-field gravity is concerned~\citep[see in particular][]{psaltis14}.

\subsubsection{Ray tracing using an analytical or numerical Kerr metric}

\begin{figure}
	\centering
	\includegraphics[height=5cm,width=12cm]{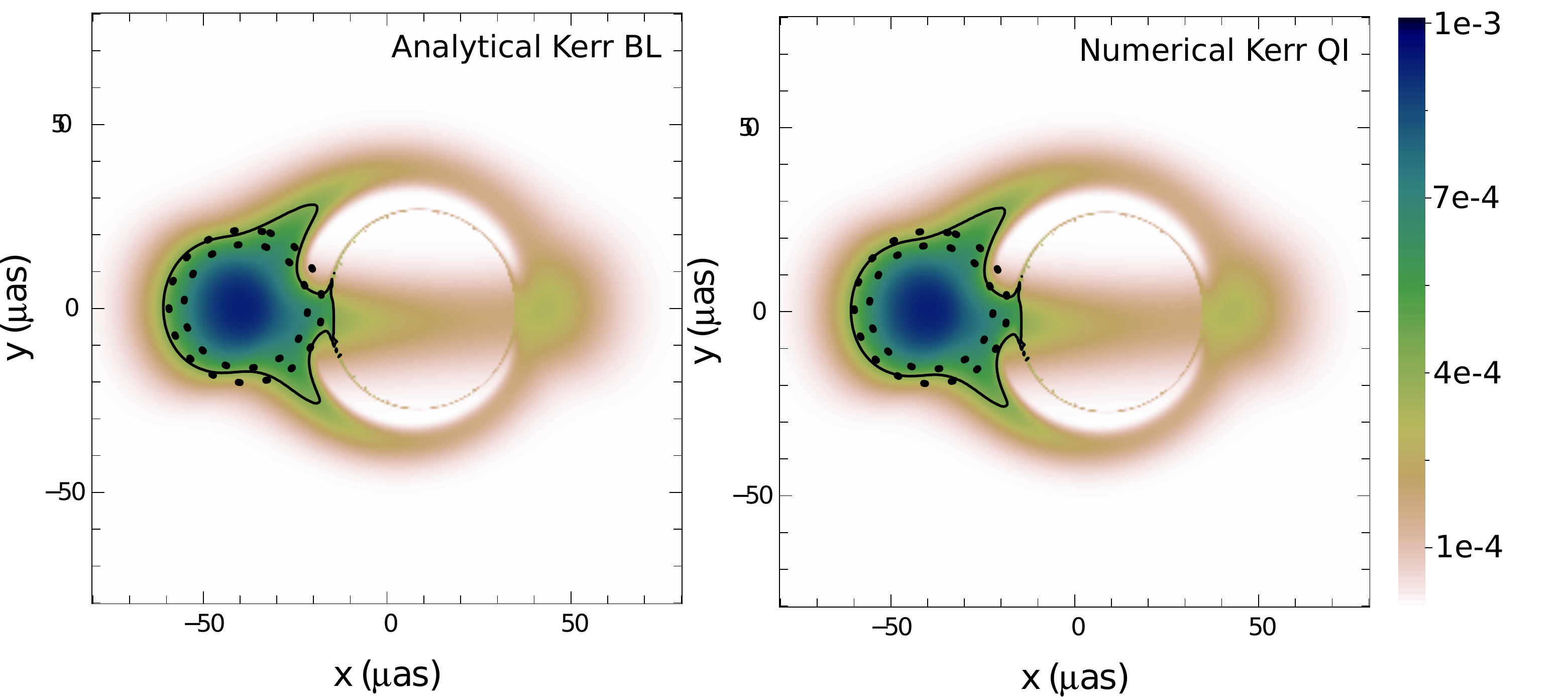} \hspace{0.5cm}
	\caption{\textit{Integration in numerical spacetimes.} \textbf{Left:} Kerr "reference" image described in Section~\ref{sec:refkerr} computed
	by the GYOTO code using the usual analytic Kerr metric with spin $a=0.9$ in Boyer-Lindquist (BL) coordinates.
	\textbf{Right:} the same image computed by GYOTO using a numerical Kerr spacetime with the same spin,
	in quasi-isotropic (QI) coordinates. The two images have the same flux within a relative error of $0.02 \%$.}
	\label{fig:kerrcompare}
\end{figure}

Imaging boson stars will necessitate integrating geodesics in a numerical spacetime.
In this section we compare the accuracy of two computations of the "reference" Kerr image.
One image is integrated using the usual analytical expression of the Kerr metric in Boyer-Lindquist
coordinates with a spin $a=0.9$. The second image is integrated in a Kerr numerical spacetime (with $a=0.9$ as well) 
computed using the LORENE
library\footnote{Available at \url{http://www.lorene.obspm.fr}}.% upon which KADATH was developed.
%KADATH and LORENE have very similar 
This spacetime is described in quasi-isotropic coordinates which differ from Boyer-Lindquist
coordinates and will be used to describe all boson-star spacetimes. 
Figure~\ref{fig:kerrcompare} shows the same strong-field image as already
illustrated in the left panel of Figure~\ref{fig:kerrref} computed by the GYOTO code in both these spacetimes.
These two images are indistinguishable by eye, and their respective fluxes differ by no
more than $0.02 \%$ demonstrating that GYOTO is able to very accurately integrate geodesics in numerical spacetimes.
We insist on the fact that the analytical and numerical spacetimes are described in very different
coordinate systems (for instance the radial coordinate values at the horizon, $r_{\mathrm{BL}}$ and $r_{\mathrm{QI}}$
for Boyer-Lindquist and quasi-isotropic coordinates, differ by a factor
$\approx 4.6$) and that the observable, Figure~\ref{fig:kerrcompare},
is the same as it should.

\subsection{Accretion tori around a boson star}
\label{sec:BS}

\subsubsection{Tori setups}

Accretion tori surrounding boson stars can be computed relatively easily,
in much the same way as in the more standard Kerr case. Our recent analysis~\citep{meliani15}
highlights some of the main properties of these structures. In this section, we are interested in
examining the modification on strong-field images imposed by the change of spacetime. As a
consequence, we will keep fixed nearly all model parameters to the values given in Table~\ref{tab:kerrbf}.
Fixing the inner radius fixes the radial extent of the torus in a given spacetime, but this radial extent 
depends quite strongly on the spacetime. Therefore, we do not decide to keep the inner radius fixed,
but rather to choose the inner radius in order to get, for all spacetimes considered, a radial extent
of roughly $20\,M$ in Boyer-Lindquist coordinates 
(which is the radial extent of the reference torus in the Kerr metric described in Section~\ref{sec:refkerr}).

Table~\ref{tab:BS} gives the parameters used for all boson-star setups. All the parameters which
are not mentioned have the same value as in Table~\ref{tab:kerrbf}. Note that the dimensionless
spin parameter $a$ can become bigger than $1$ as opposed to the Kerr black hole case. There is nothing
particular with a boson-star spacetime with a spin bigger than $1$. In particular there is of course
no naked singularity (as would be the case in the Kerr spacetime with $a>1$).
\begin{table}[h!]
\footnotesize
\centering                          % used for centering table
\begin{tabular}{c c c c }
\\
\hline \hline
\\
$(k,\omega)$	  & $M$ & $a$ & $r_{\mathrm{in}}$ 
\\
\hline \hline
\\
$(0,0.83 \, m/\hbar)$    &   $0.63\,\mathcal{M}$ &  $0.00$    &    $4.39\,{M} $         \\
$(0,0.9 \, m/\hbar)$    &   $0.60\,\mathcal{M}$ &  $0.00$    &    $5.80\,{M} $         \\
$(1,0.7\,m/\hbar)$    &   $1.26\,\mathcal{M}$ &  $0.82$ &    $2.72\,{M}$  \\
$(1,0.8\,m/\hbar)$    &   $1.31\,\mathcal{M}$ &  $0.80$     &    $2.84\,{M} $ \\
$(1,0.9\,m/\hbar)$    &   $1.12\,\mathcal{M}$ &  $0.92$     &    $4.90\,{M}$  \\
$(4,0.7\,m/\hbar)$    &   $3.90\,\mathcal{M}$ &  $1.13$ &    $3.34\,{M}$  \\
$(4,0.8\,m/\hbar)$    &   $3.35\,\mathcal{M}$ &  $1.27$ &    $2.92\,{M}$  \\
$(4,0.9\,m/\hbar)$    &   $2.52\,\mathcal{M}$ &  $1.64$ &    $5.30\,{M}$  \\
\hline        \hline                                   %inserts single line
\end{tabular}
\caption{\textit{Parameters used to compute accretion tori in the various boson-star spacetimes considered here.
$M$ is the ADM mass given in units of $\mathcal{M}=m_p^2 / m$. }}
\label{tab:BS}
\end{table}

\begin{figure}
	\centering
	\includegraphics[height=5cm,width=5cm]{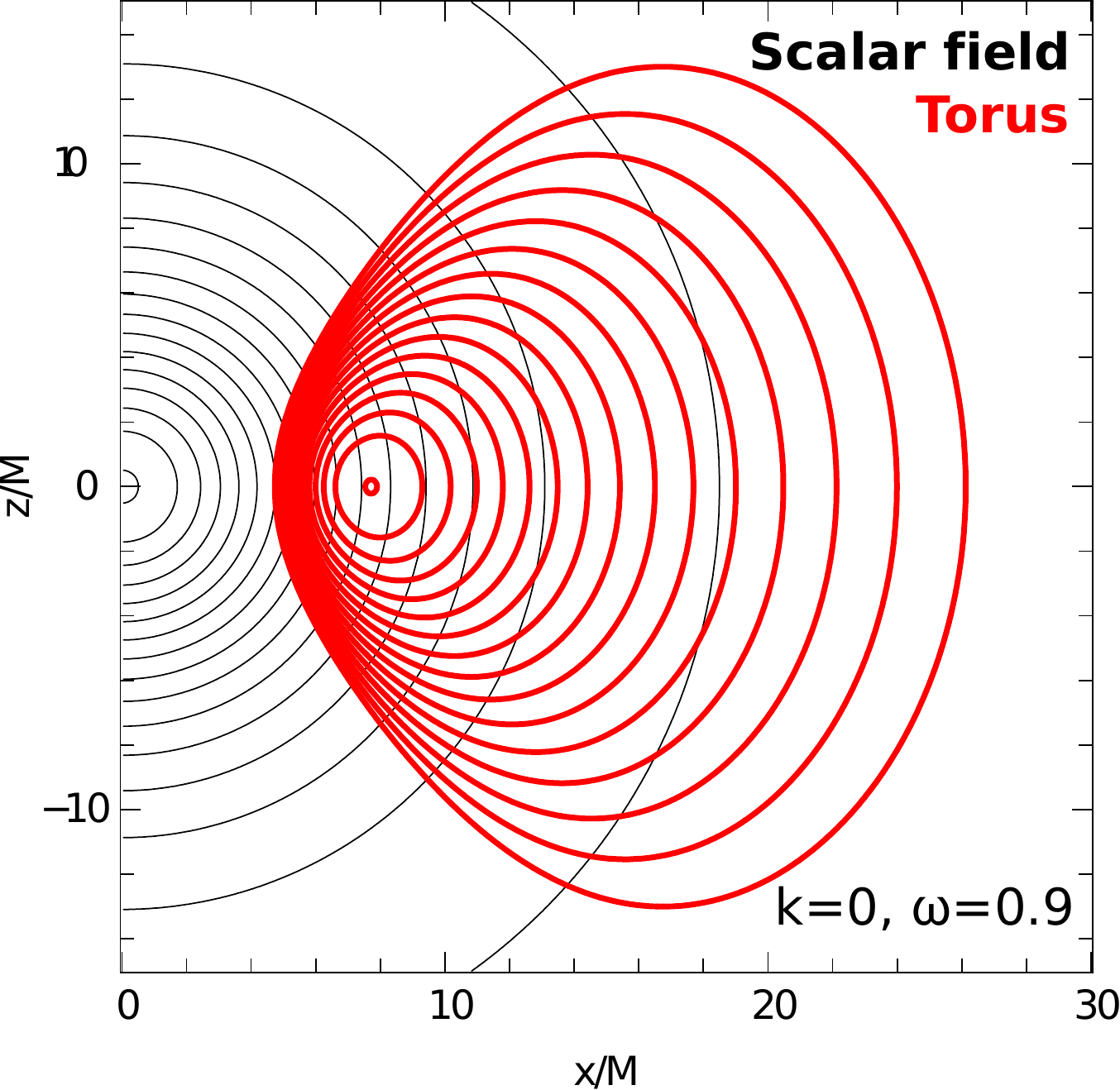} 
	\includegraphics[height=5cm,width=5cm]{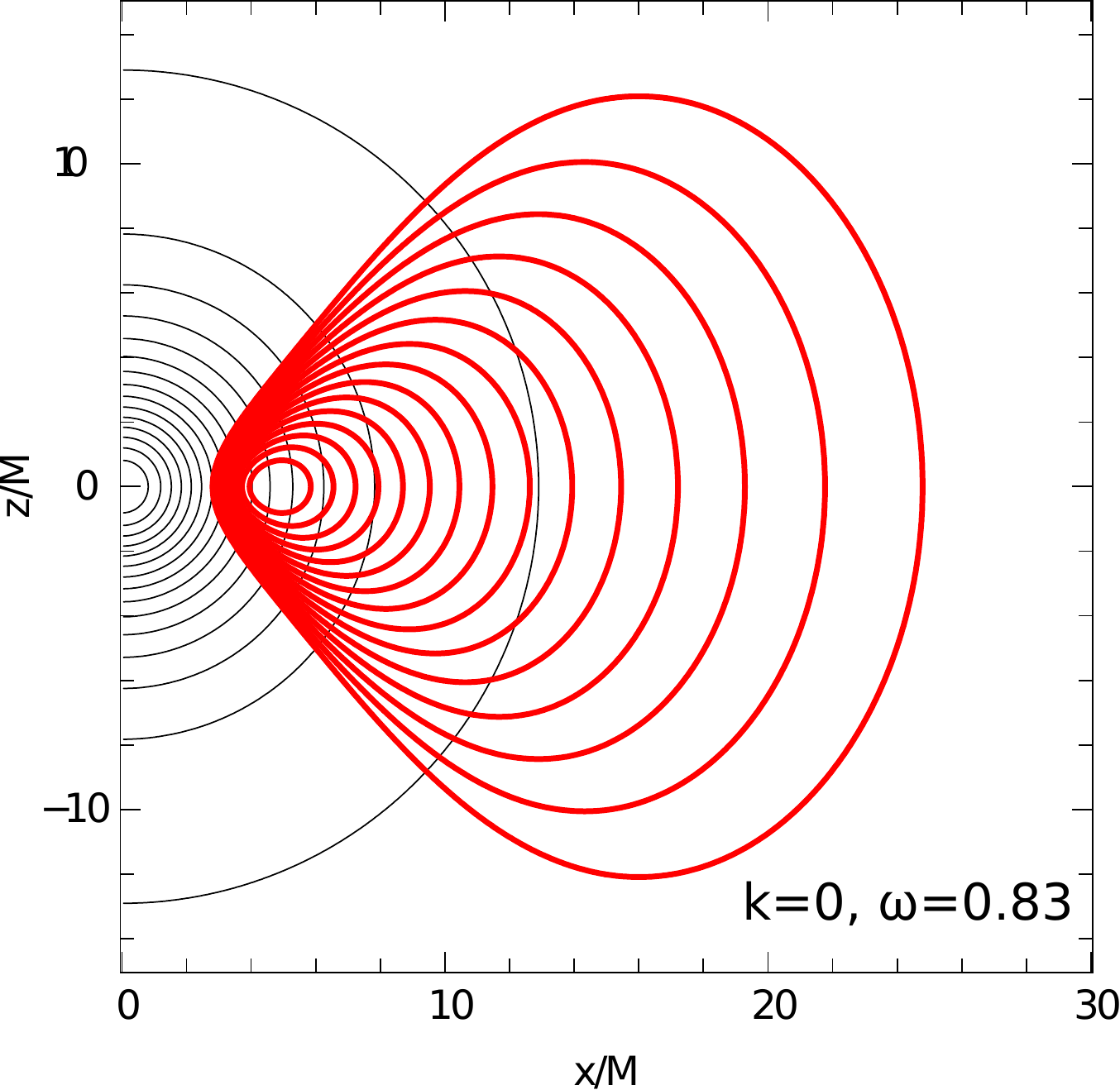}\\
	\includegraphics[height=5cm,width=5cm]{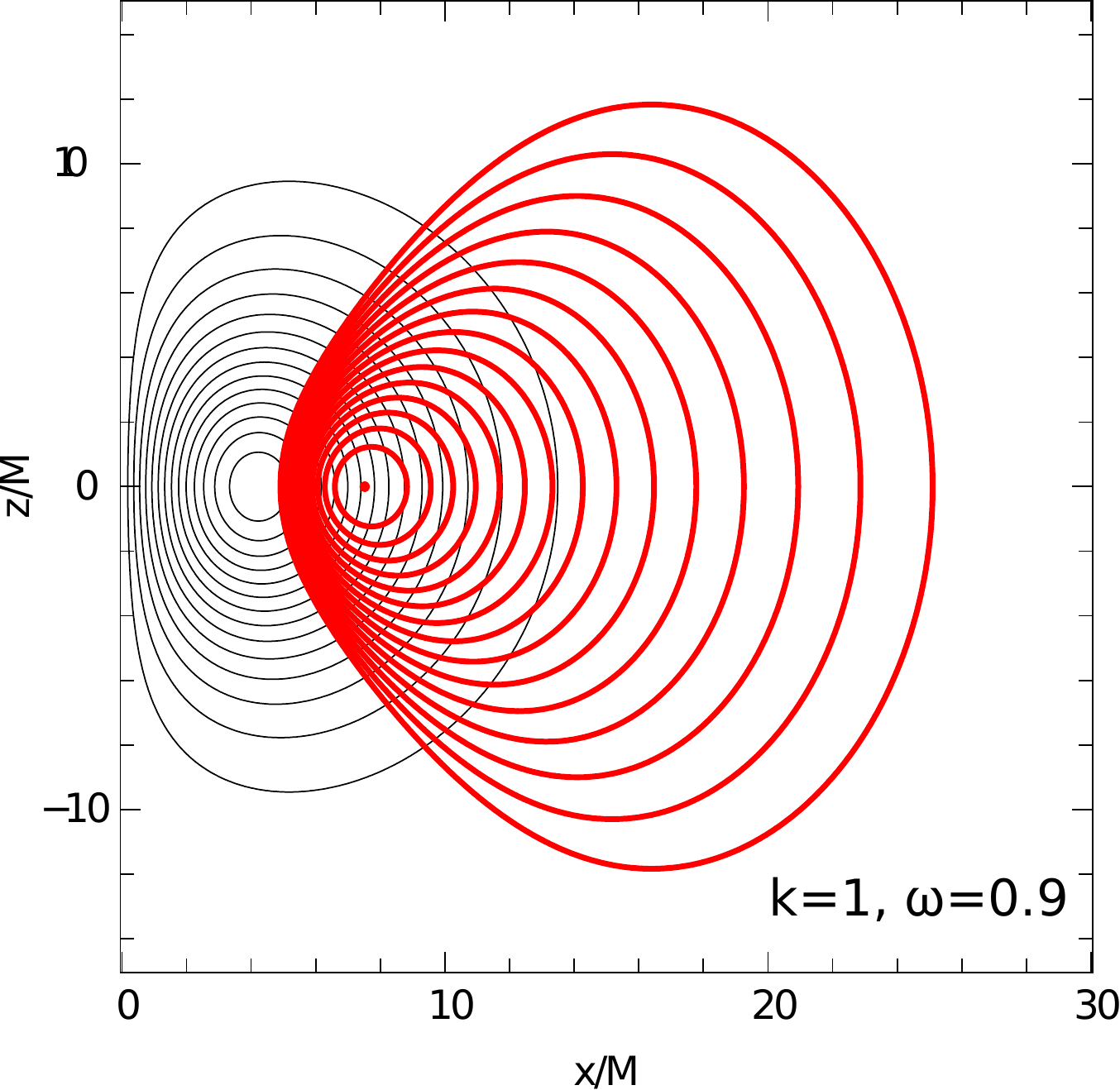}
	\includegraphics[height=5cm,width=5cm]{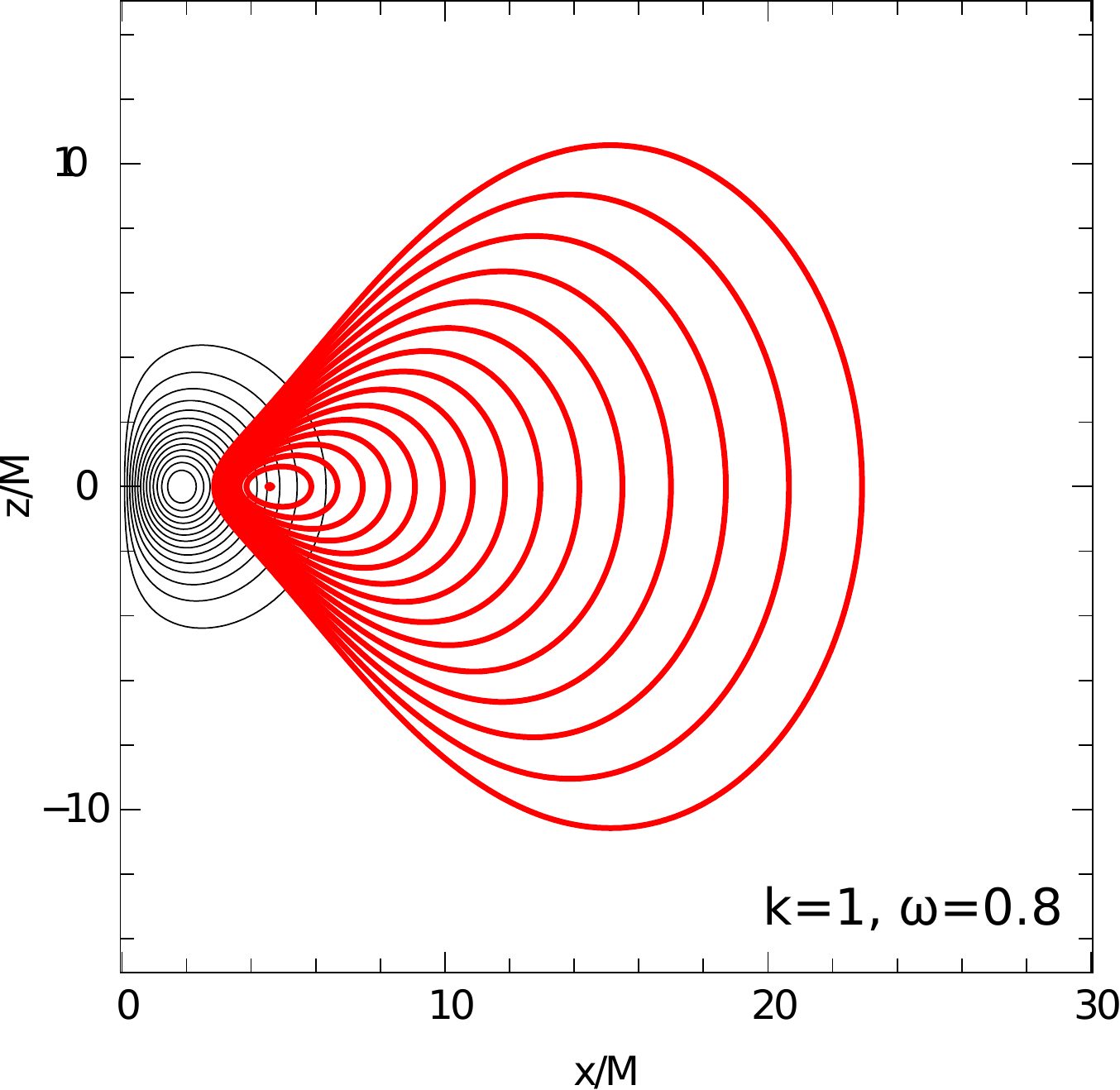}
	\includegraphics[height=5cm,width=5cm]{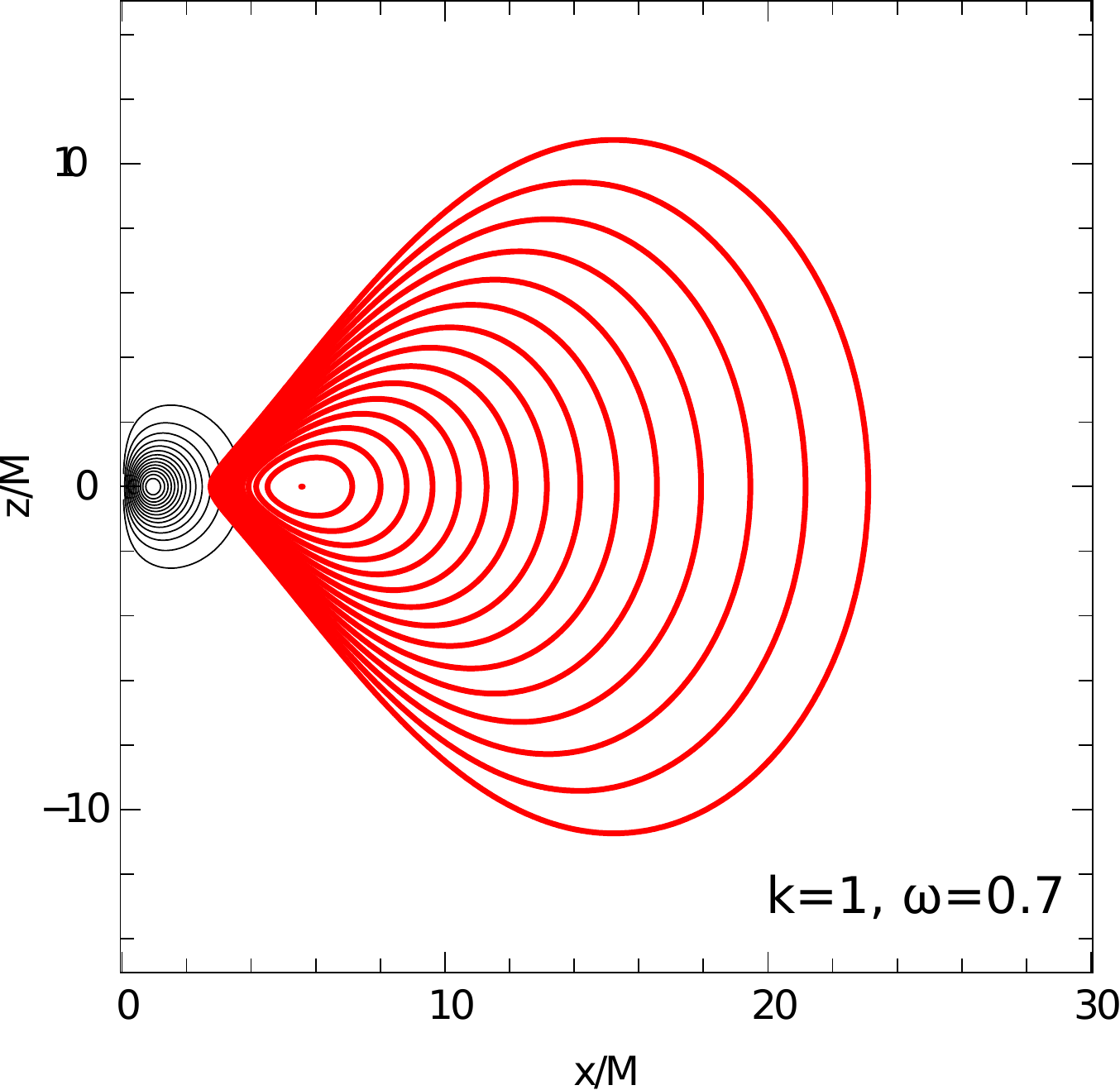} \\
	\includegraphics[height=5cm,width=5cm]{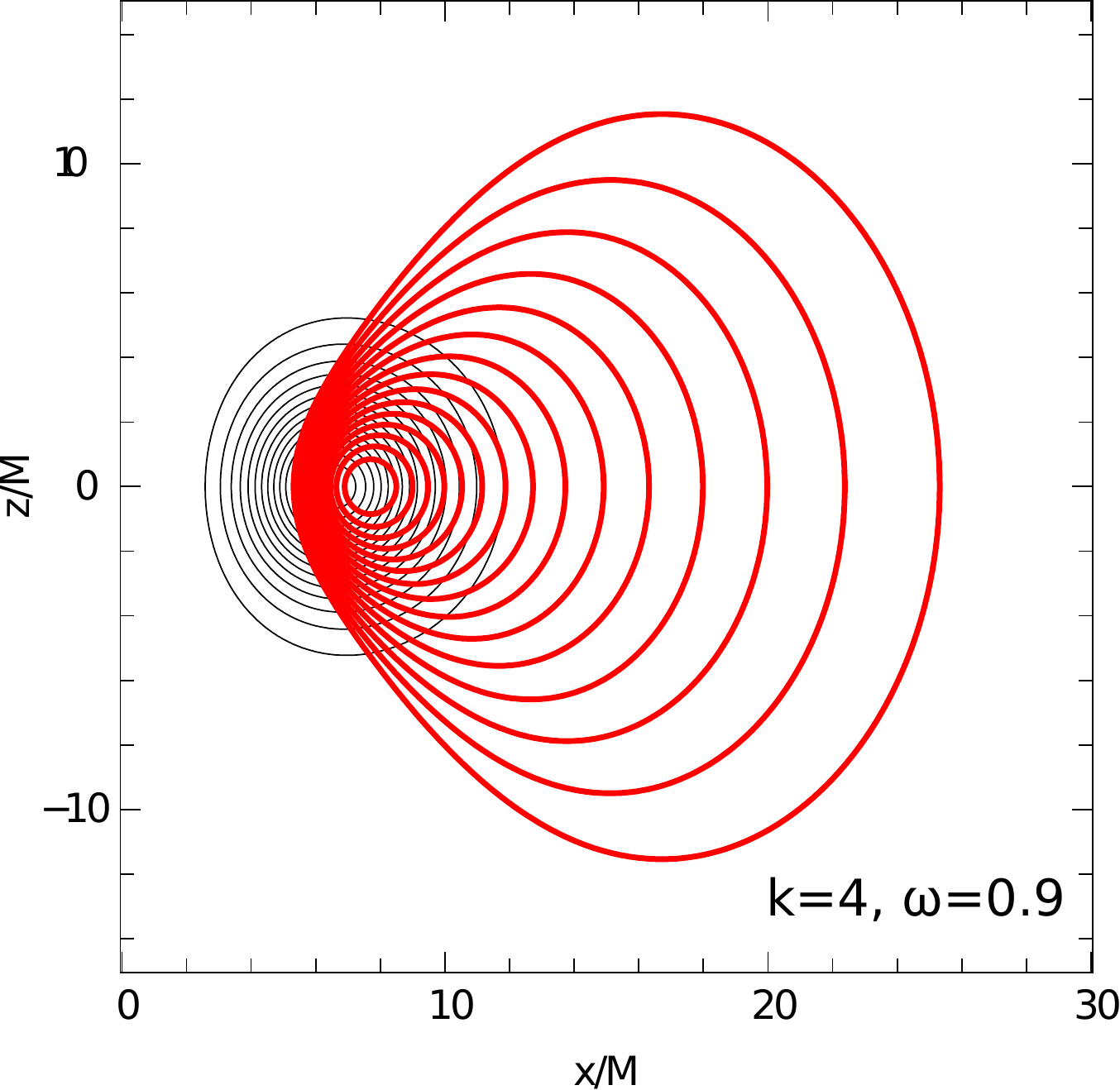}
	\includegraphics[height=5cm,width=5cm]{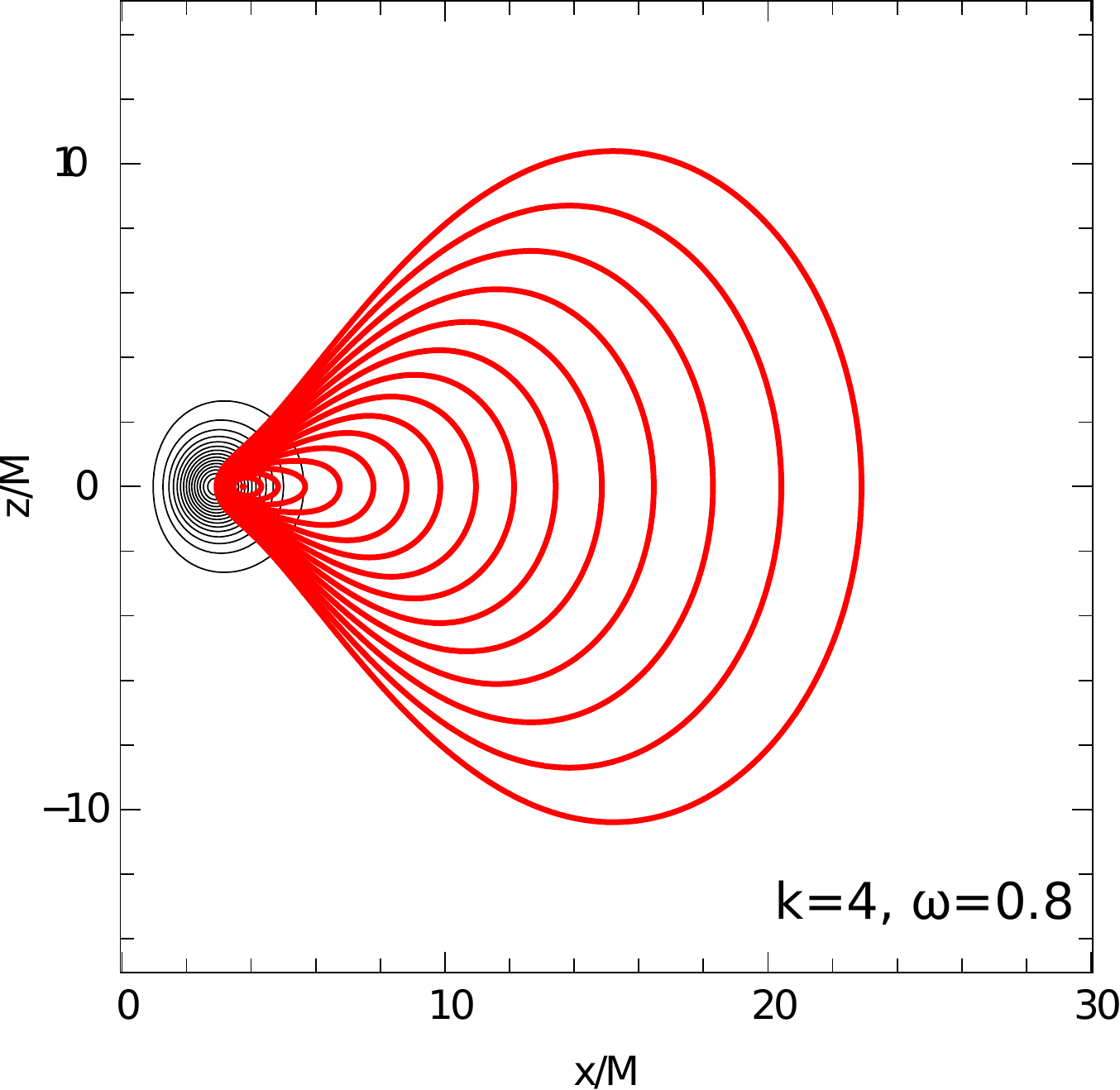}
	\includegraphics[height=5cm,width=5cm]{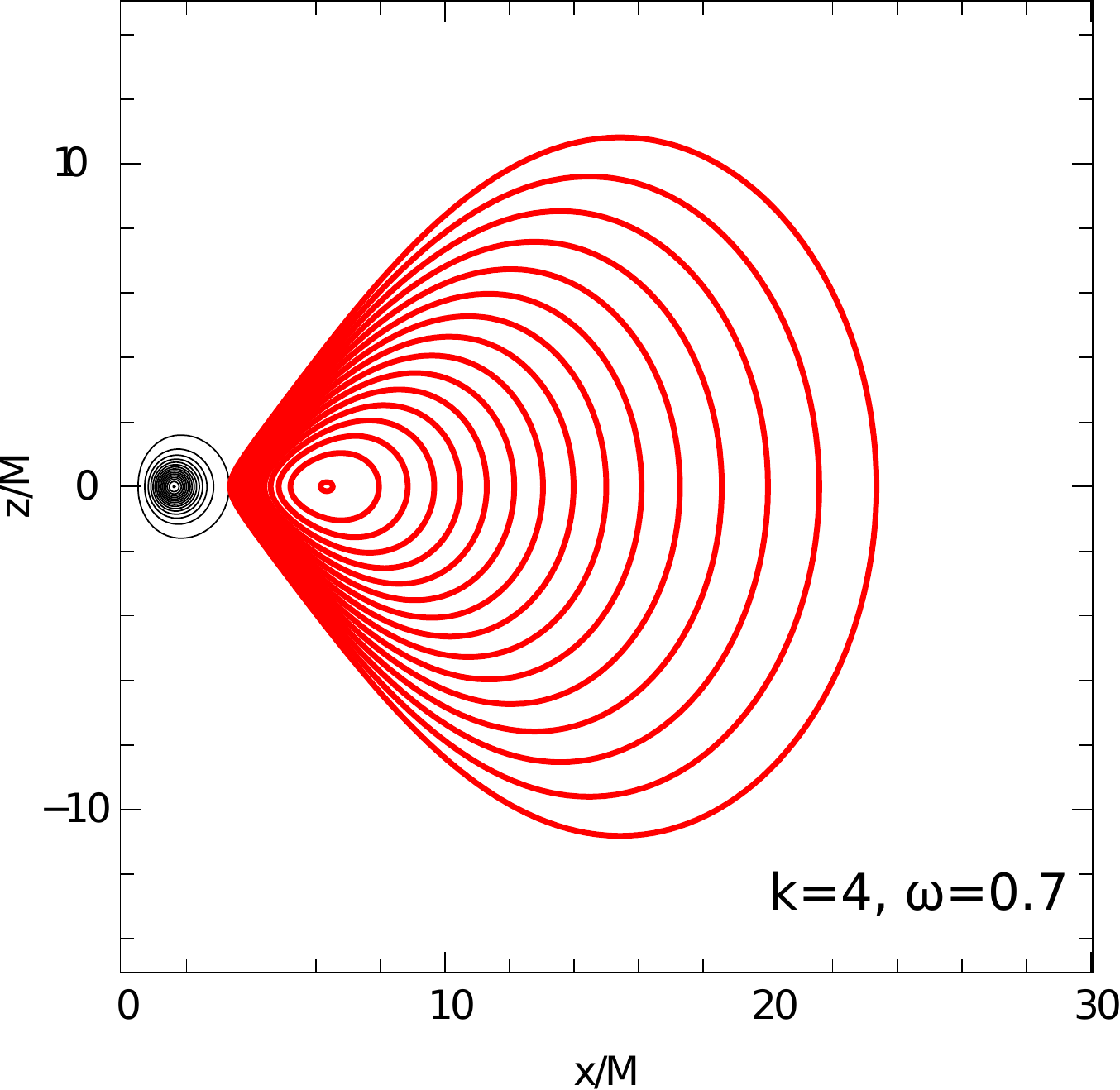}
	\caption{\textit{Boson-star accretion tori contours.} Equi-pressure contours of the accretion torus (red) and contours of the
	scalar field distribution (black) for the various setups described in Table~\ref{tab:BS},
	in the $(x,z)$ plane, where $z$ is a coordinate along the rotation axis.
	The axes are labeled in units of the boson star total mass $M$.
	Boson star rotation is increasing from top to bottom (towards higher $k$), 
	and the spacetime is more and more
	relativistic from left to right (towards smaller $\omega$). }
	\label{fig:contours}
\end{figure}

Figure~\ref{fig:contours} shows the contours of the equi-pressure surfaces of these
tori together with the contours of the scalar field modulus $\phi$. These panels highlight
the fact that when the boson star is rotating, the scalar field distribution has a toroidal
topology. The name boson "star" (suggestive of a spherical topology) is thus misleading for such objects, however we
keep using it for historical reasons. While the contours of the torus remain rather similar
for all spacetimes (including the Kerr spacetime, see the right panel of Figure~\ref{fig:kerrref}), the scalar field distribution is a bit more peaked for higher rotation
and much more peaked for more relativistic spacetimes. This will translate in more
important lensing effects in the strong-field region for very relativistic boson stars.
We note also that the accretion torus and the
scalar field distribution overlap in regions where the scalar field is still far from $0$ (i.e., 
rather close to the center of the distribution), 
which does not lead to any physical effect as we
assume that there is no interaction between normal (baryonic) matter and the scalar field.

\subsubsection{Images and spectra}

\begin{figure}
	\centering
	\includegraphics[height=5cm,width=5cm]{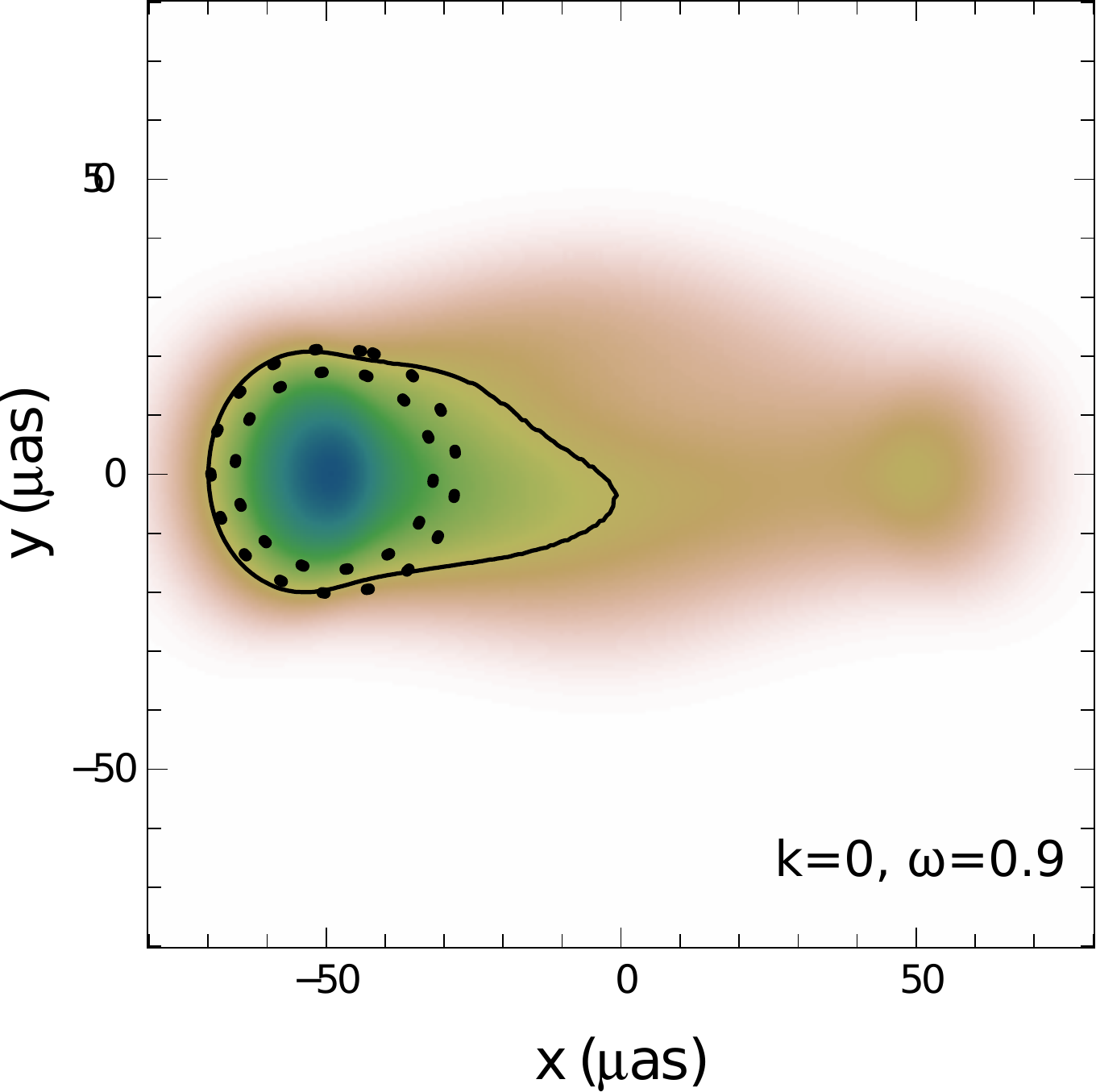} 
	\includegraphics[height=5cm,width=6cm]{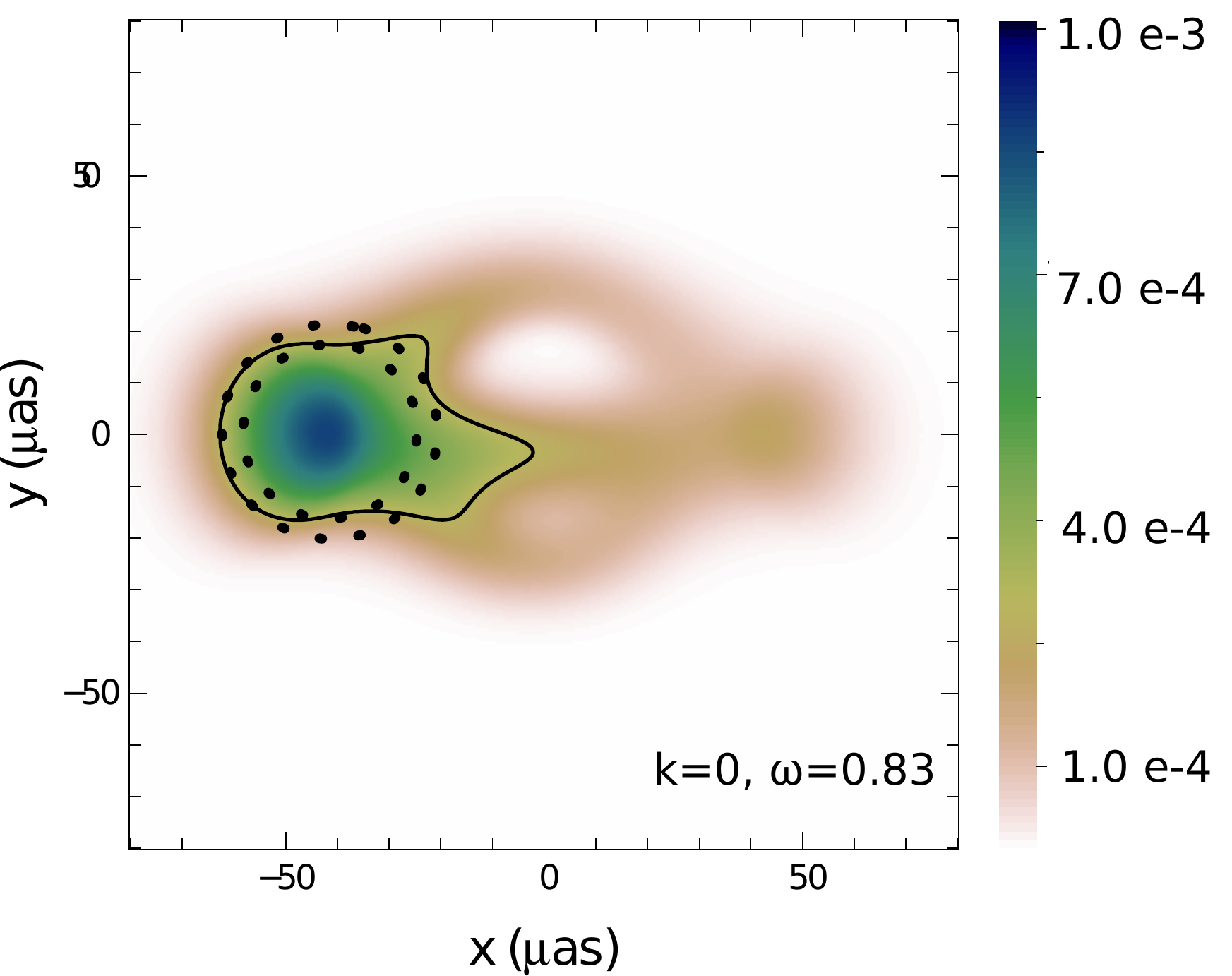} \\
	\includegraphics[height=5cm,width=5cm]{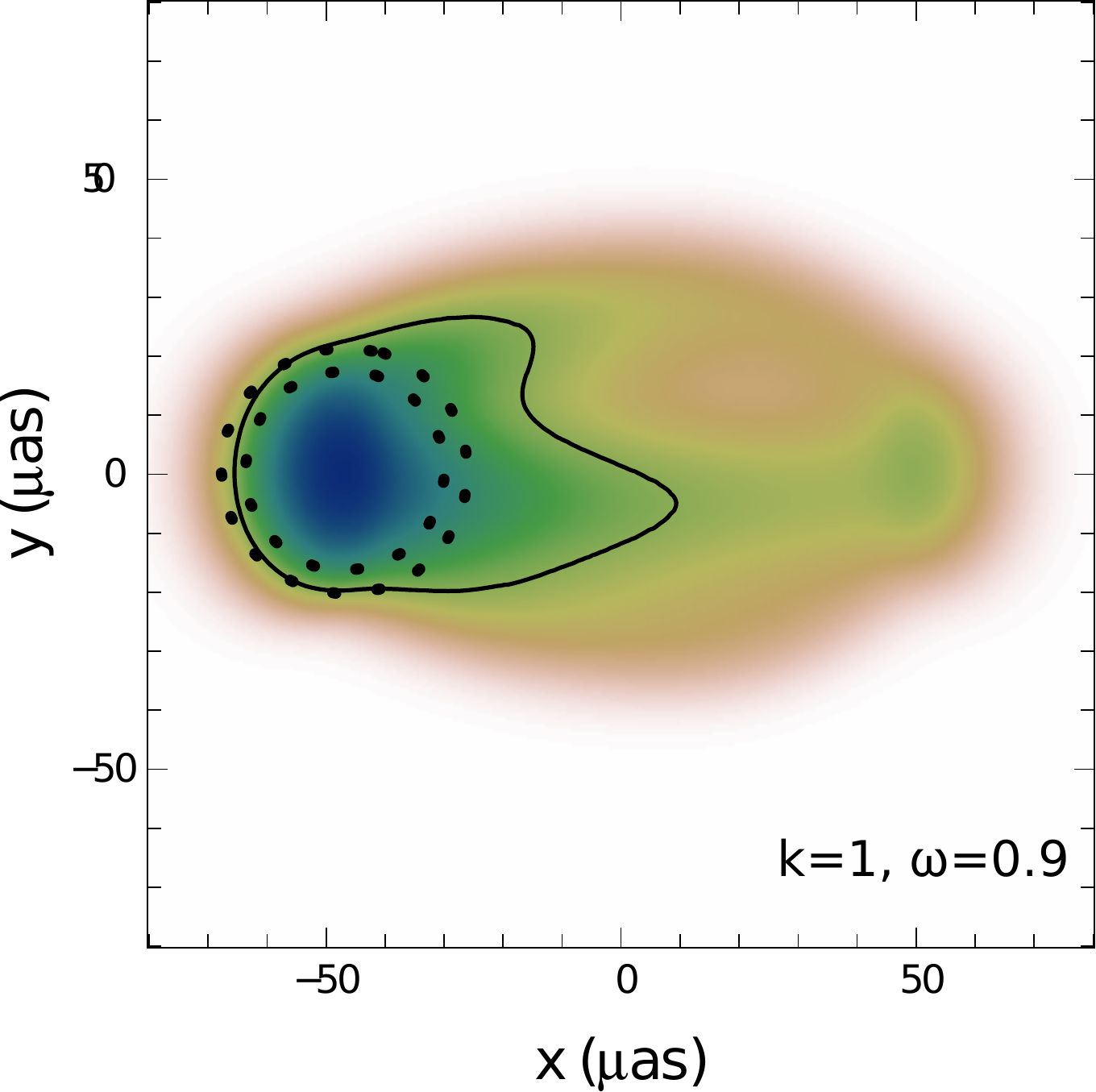}
	\includegraphics[height=5cm,width=5cm]{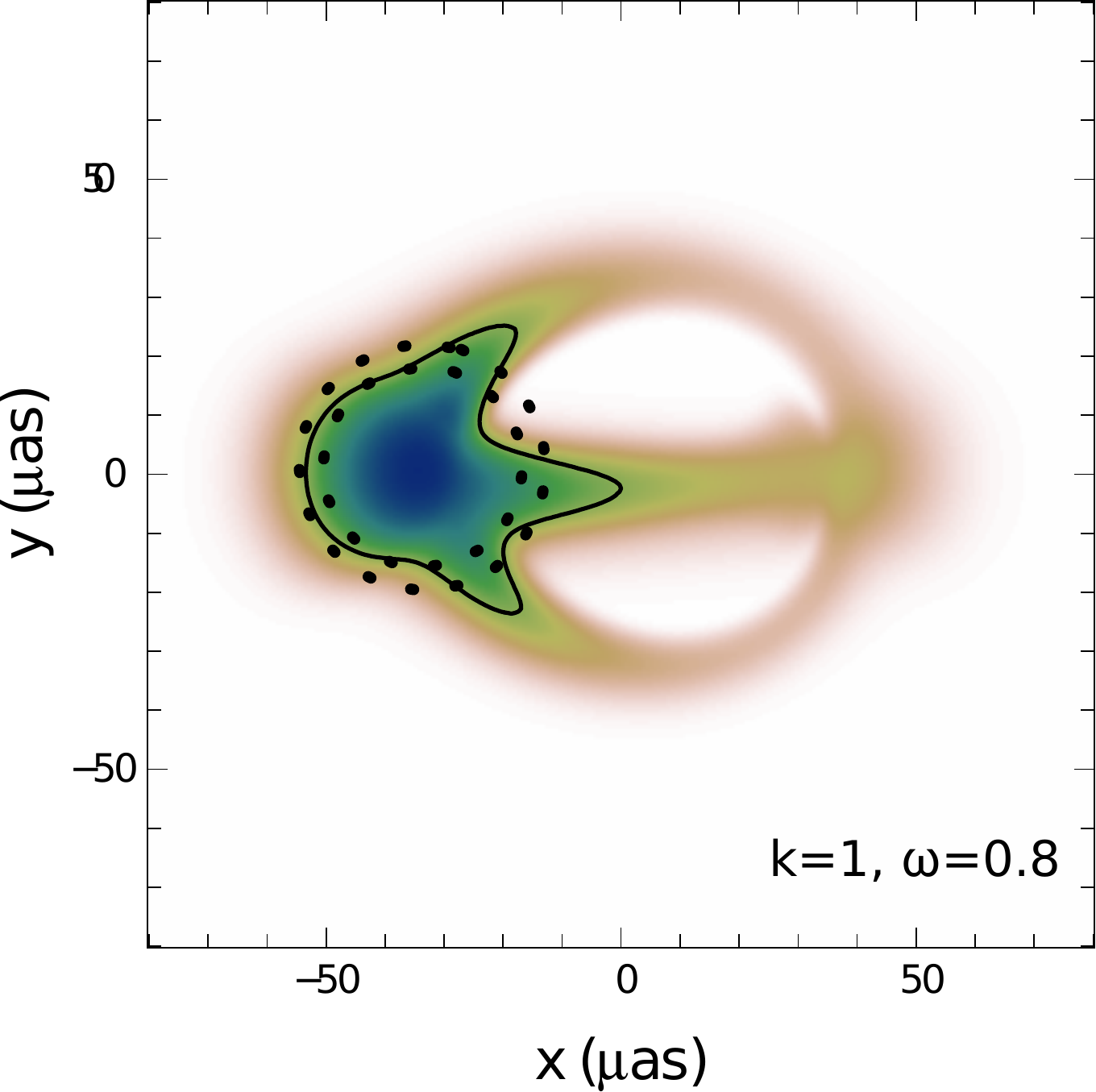}
	\includegraphics[height=5cm,width=5cm]{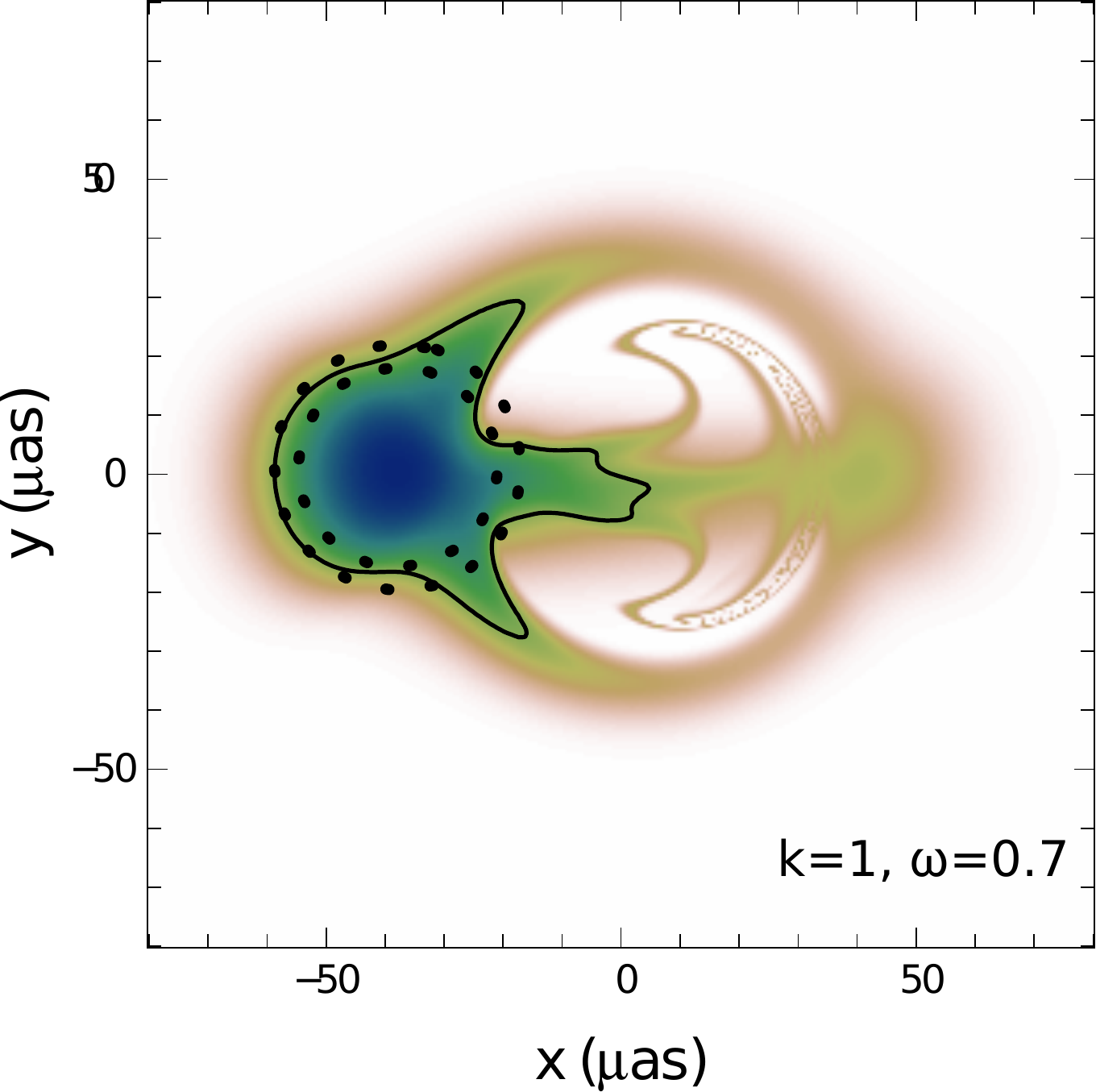} \\
	\includegraphics[height=5cm,width=5cm]{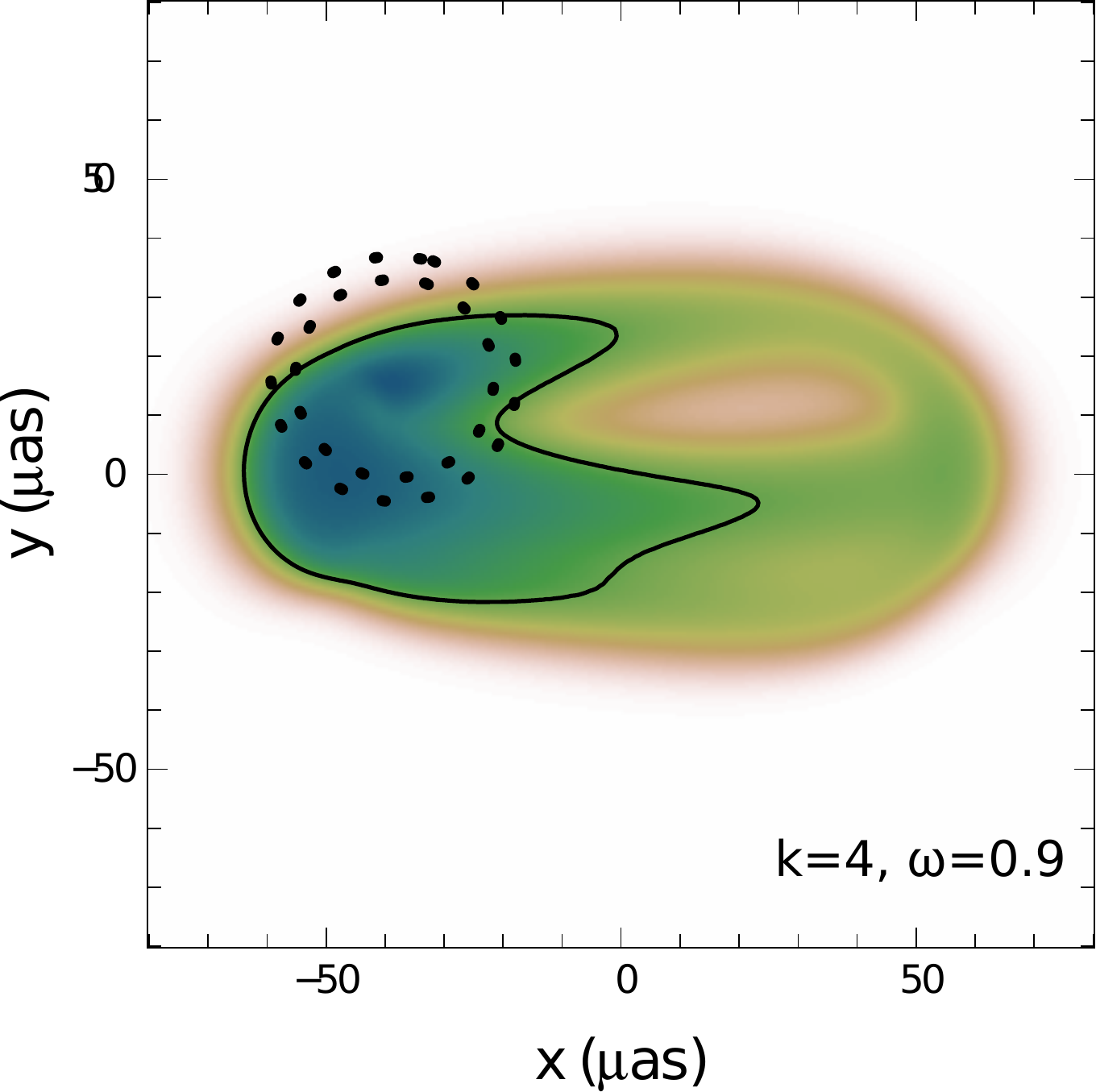}
	\includegraphics[height=5cm,width=5cm]{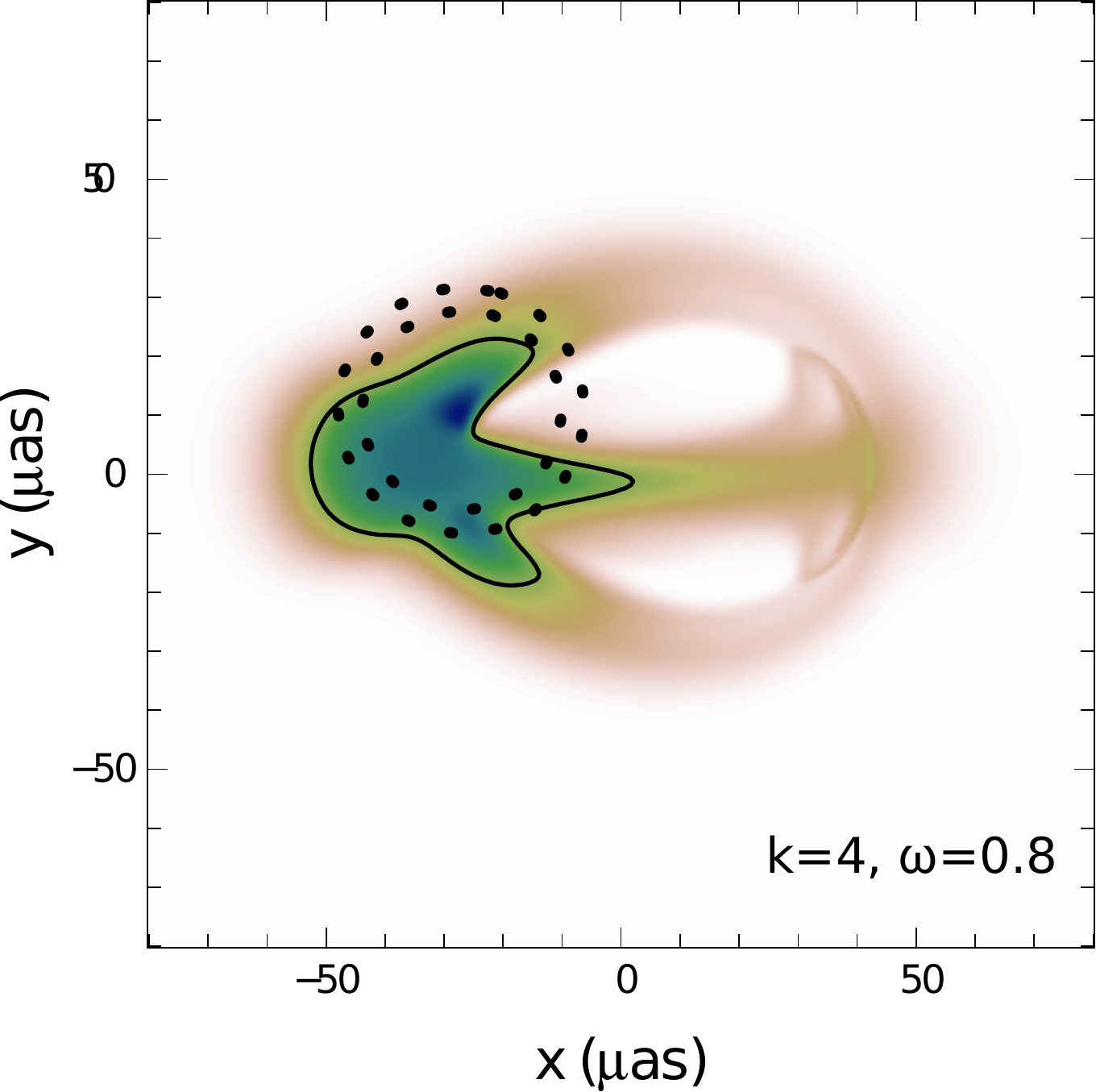}
	\includegraphics[height=5cm,width=5cm]{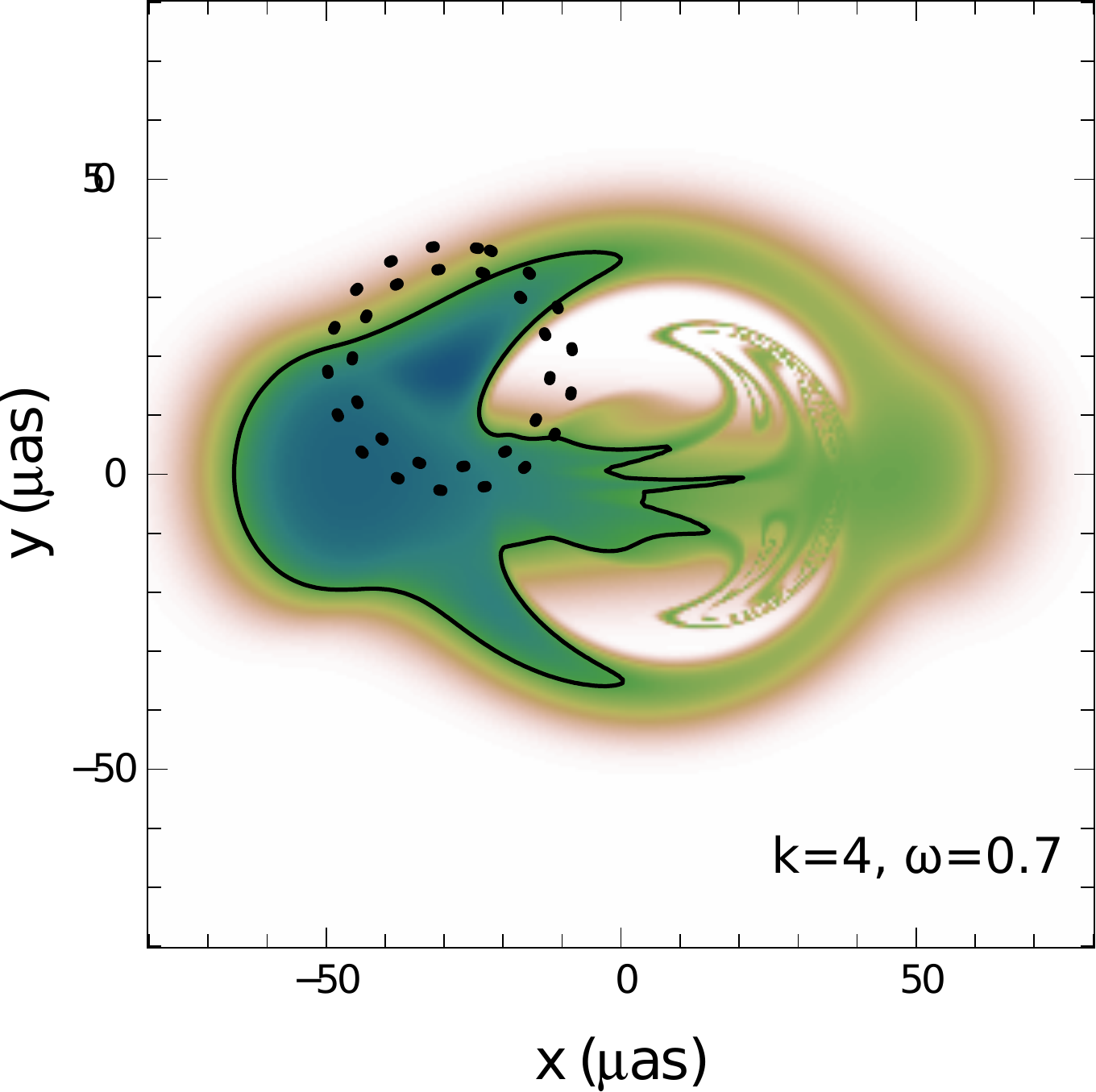}
	\caption{\textit{Boson-star images.} Maps of specific intensity distribution for the various boson-star setups given in Table~\ref{tab:BS}.
	The color bar at the top right is valid for all panels and is graduated in cgs units.
	The dotted circles show the $1\sigma$ confidence limit on the angular size of the emitting region
	imposed by the first VLBI measurements~\citep{doeleman08}. They are centered on the maximum of the intensity distribution.
	The solid black contour encompasses the region emitting $50\%$ of the total flux.
	The axes are labeled in $\mu$as, as measured on the distant observer's screen.
	Boson star rotation is increasing from top to bottom (towards higher $k$), 
	and the spacetime is more and more
	relativistic from left to right (towards smaller $\omega$). }
	\label{fig:images}
\end{figure}

Figure~\ref{fig:images} shows the $1.3$~mm images of all the tori setups surrounding
boson stars given in Table~\ref{tab:BS}. For less-relativistic setups ($\omega=0.9\,m/\hbar$), which
are closer to empty space (remember that $\omega=m/\hbar$ corresponds to empty space,
see Section~\ref{sec:BSintro}),
images show a smooth distribution of specific intensity for all values of $k$, with no strong
gradient (no "hole" at the center of these images). This is close to the image one would get in a Newtonian spacetime
of a thick torus seen edge-on. For all values of $k$ also, a region with much lower intensity
value (a "hole") appears in the image as $\omega$ decreases. For very relativistic spacetimes
($\omega=0.7$) this "hole" is accompanied by a bow-shape structure, which is the equivalent of
the Kerr photon ring.

\begin{figure}
	\centering
	\includegraphics[height=6cm,width=6cm]{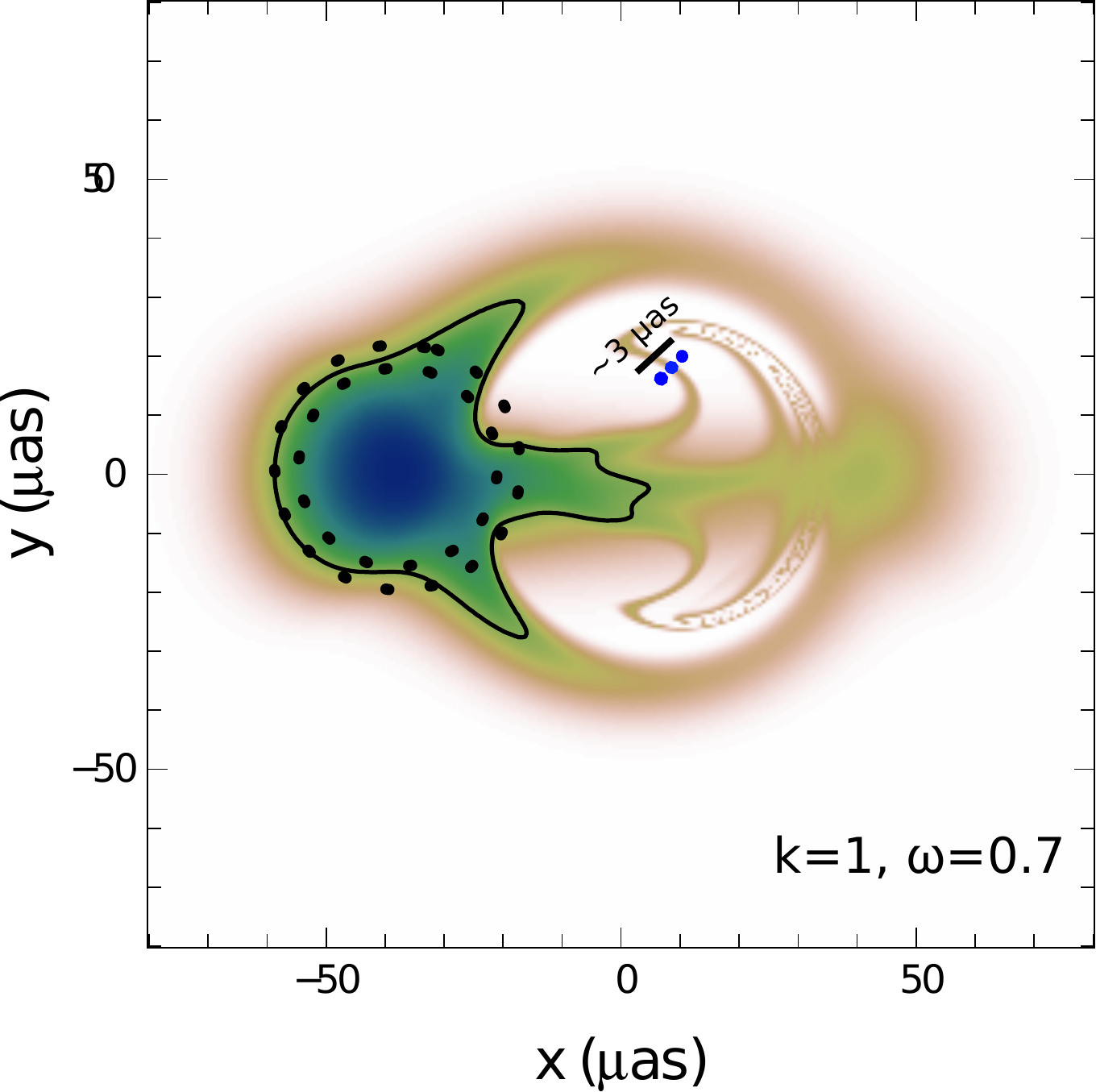} \\
	\includegraphics[height=5cm,width=15cm]{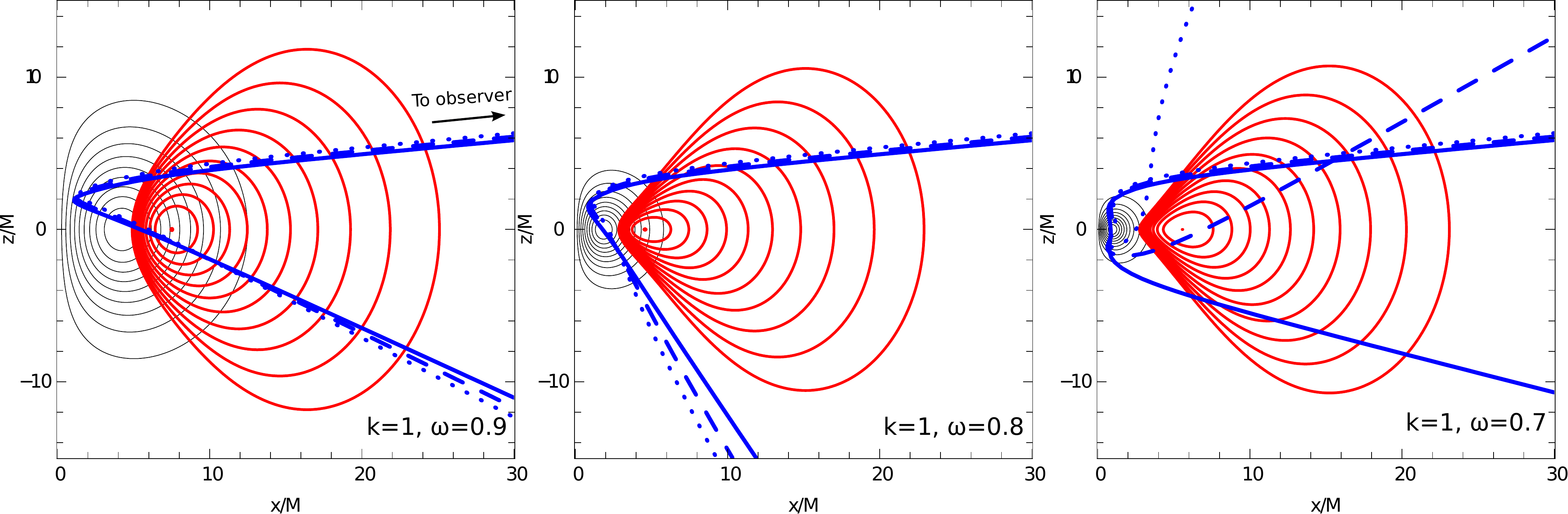} \\
	\includegraphics[height=5cm,width=5cm]{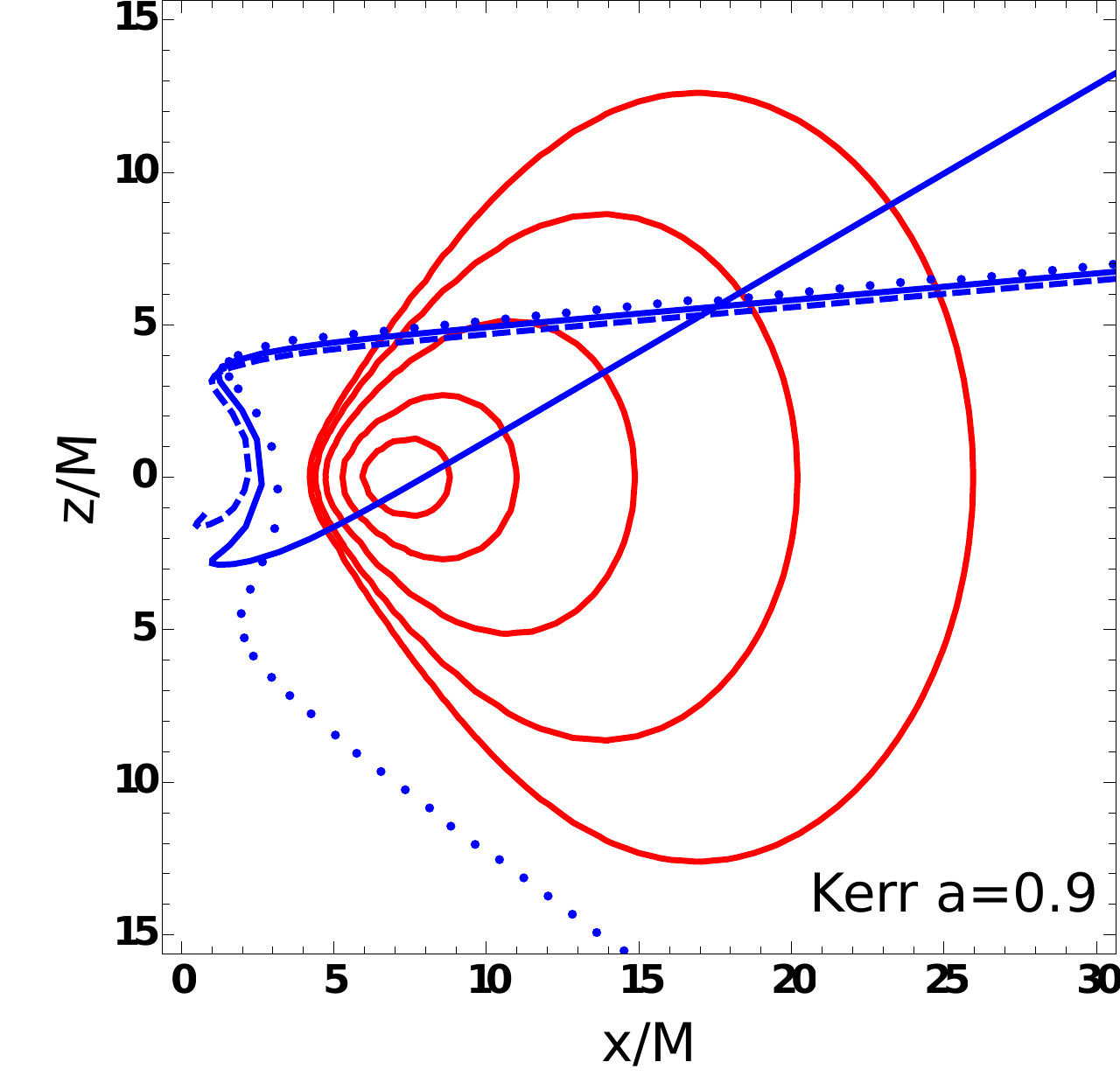}
	\caption{\textit{Light bending.} \textbf{Top row}: the $k=1$, $\omega=0.7\,m/\hbar$ image with $3$ blue dots corresponding
	to the directions on sky of the $3$ geodesics represented in the middle row panels. These $3$ directions
	are separated by only $\approx 3\, \mu$as.
	\textbf{Middle row: }same as Figure~\ref{fig:contours} for $k=1$ boson stars, with $3$ photon geodesics over-plotted 
	in blue in each panel, corresponding to the $3$ directions on sky highlighted in the top panel. 
	The geodesics are integrated backwards in time from the distant observer.
	They are computed in $3$ space dimensions $(r,\theta,\pp)$ and are projected here
	in $(x = r \sin \theta,z= r \cos \theta)$ whatever $\pp$.
	Mind that part of the geodesics curvature on these plots is due to the projection 
	from $3$ to $2$ space dimensions. 
	The difference of magnitude of the lensing effect depending on the value of $\omega$ appears clearly.
	\textbf{Bottom row:} the same for the reference Kerr $a=0.9$ case. The $3$ geodesics represented here do not correspond to the
	same directions on sky as the previous ones. They are associated to the vicinity of the Kerr photon ring, i.e. to the most lensed
	geodesics in the Kerr spacetime. Note that the dashed geodesic asymptotically approaches the event horizon.}
	\label{fig:geods}
\end{figure}

In order to understand these transitions, Figure~\ref{fig:geods} (top row) shows again the torus pressure and scalar field contours
for $k=1$ spacetimes,
together with $3$ geodesics projected in the $(x = r \sin \theta,z= r \cos \theta)$ plane. These panels show the increasing gravitational lensing
effects on null geodesics as $\omega$ decreases and the spacetime becomes more relativistic. When the lensing effect
is strong enough, a low-intensity region appears at the center of the images. When this effect is even stronger, two
geodesics corresponding to very similar directions on sky (within $\approx 1\,\mu$as) can have very different trajectories, leading to the development of the bow-shape structure. The bottom panel
of Figure~\ref{fig:geods} shows that this is similar to what causes the appearance of the Kerr photon ring.
This bow-shape structure characteristic of very relativistic boson stars was first highlighted very recently by~\citet{cunha15}.
Their Figure~4, middle-right panel shows an extremely similar structure to our $k=1$, $\omega=0.7$ image.
This structure has a very comparable angular size to that of the reference Kerr photon ring. The most distant part of the
Kerr photon ring from the center of coordinates (to the right of the image) is located at $\approx35.5\,\mu$as while
the most distant part of the bow-shape structure for the $(k=1, \omega=0.7\,m/\hbar)$ spacetime is located
at $\approx34.5\,\mu$as from the center of coordinates.

The superposition of the low-flux central region and of this bow-shape structure is extremely similar
to the shadow+photon ring familiar structure in the Kerr spacetime. In particular, it shows that detecting
a shadow (i.e. a low-flux region surounded by a bright portion of arc) is not sufficient to tell the existence of an
event horizon, as suggested by~\citet{falcke00}. It is probable that after distortion by the instrument's
response function, it would be impossible to differentiate a Kerr image from a very relativistic boson-star image.

\begin{figure}
	\centering
	\includegraphics[height=6cm,width=12cm]{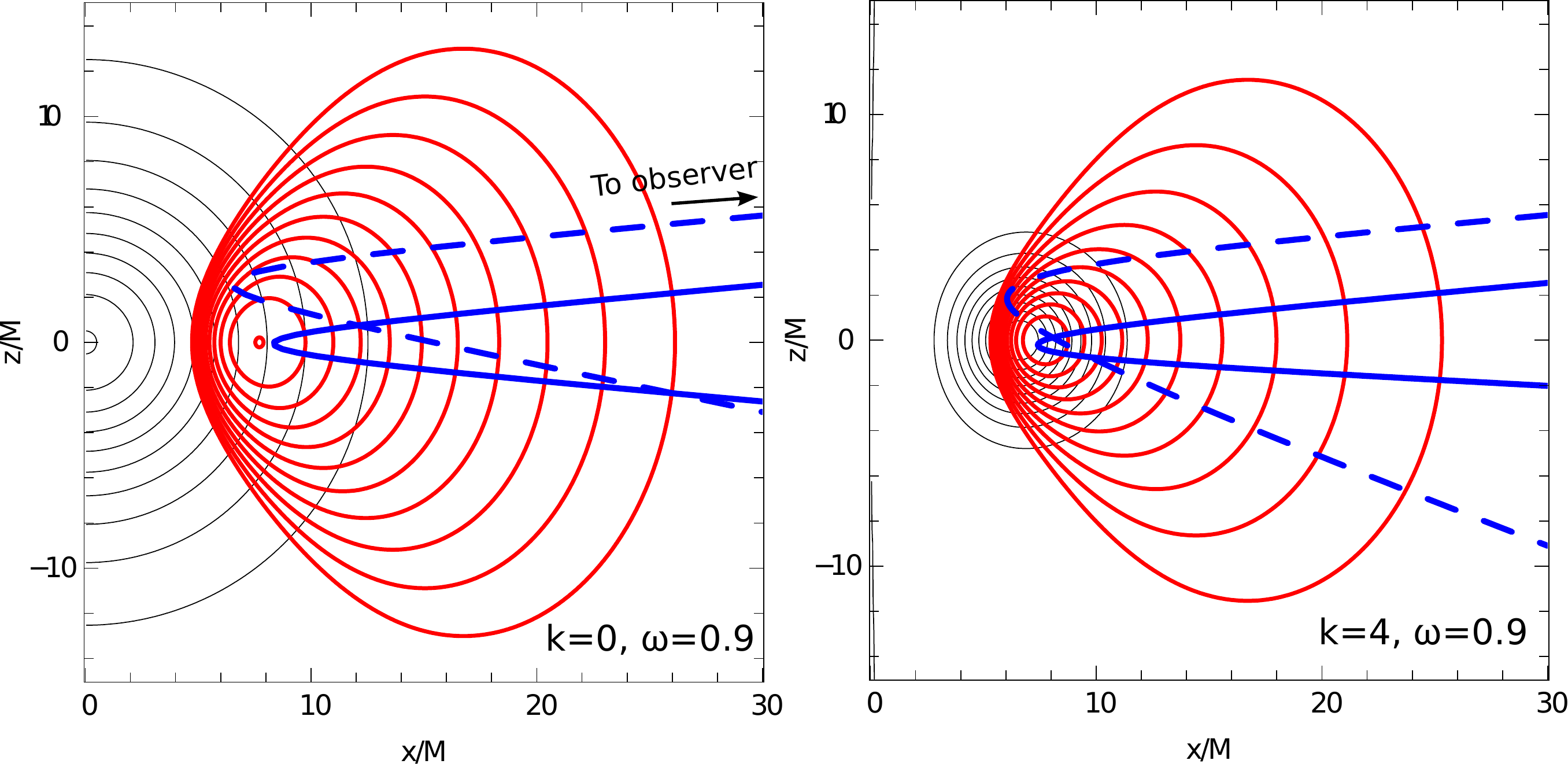}
	\caption{\textit{Intensity maximum location.} Contours of the $\omega=0.9\,m/\hbar$, $k=0$ (left) and $k=4$ (right) spacetimes.
	In blue, two geodesics are over-plotted. The solid one corresponds to the direction on the sky of the maximum of
	the intensity distribution of the $(k=0,\omega=0.9\,m/\hbar)$ setup (see Figure~\ref{fig:images}, upper left panel).
	The dashed one corresponds to the direction on the sky of the maximum of
	the intensity distribution of the $(k=4,\omega=0.9\,m/\hbar)$ setup (see Figure~\ref{fig:images}, lower left panel).
	}
	\label{fig:maximum}
\end{figure}

We note here a particularity of the $k=4$ images. All other spacetimes give rise to an intensity
distribution peaked more or less at the same point, to the left of the image in our geometry.
This location corresponds to the maximum of the relativistic beaming effect due to the
enhancement of radiation when the emitter is traveling towards the observer. However,
the maximum of the intensity distribution is somewhat shifted with respect to this maximum
beaming location for all $k=4$ spacetimes. This is mainly due to the
stronger bending of light rays as explained in Figure~\ref{fig:maximum}.
This Figure compares the two geodesics corresponding to the location on sky of the intensity maxima
of the $(k=0,\omega=0.9\,m/\hbar)$ and $(k=4,\omega=0.9\,m/\hbar)$ spacetimes. It shows that the
geodesic corresponding to the maximum intensity location of the $(k=4,\omega=0.9\,m/\hbar)$ spacetime
(dashed blue, right panel) visits the very central parts of the torus, which translates in a high intensity. On the contrary,
the same geodesic in the $(k=0,\omega=0.9\,m/\hbar)$ spacetime (dashed blue, left panel) always stays rather far from the innermost
torus regions. Strong light bending thus somewhat changes the flux distribution for $k=4$ spacetimes.

%We note here a particularity of the $k=4$ images. Typically a strong-field image has its maximum of specific
%intensity on the approaching side of the torus (to the left, due to the relativistic beaming effect that enhances radiation in the direction
%of the emitter's motion. For our geometry, the approaching side is to the left, and the maximum of the intensity
%maps is indeed on the left side of the distribution for all $k=0$ and $k=1$ spacetimes.
%However, for $k=4$ spacetimes, the specific intensity maximum is shifted with respect
%to the maximum beaming point. The beaming effect
%is still maximum on the left side of the $k=4$ images and redshift values are of the
%same order as in other spacetimes. 
%But the details of the integration
%of the radiative transfer equation leads to a different intensity distribution:
%this distribution is more peaked towards the maximum beaming point for $k=0$ and $k=1$ spacetimes than for
%$k=4$ (as is clear from Figure~\ref{fig:images}). 
 
\begin{figure}
	\centering
	\includegraphics[height=7cm,width=8cm]{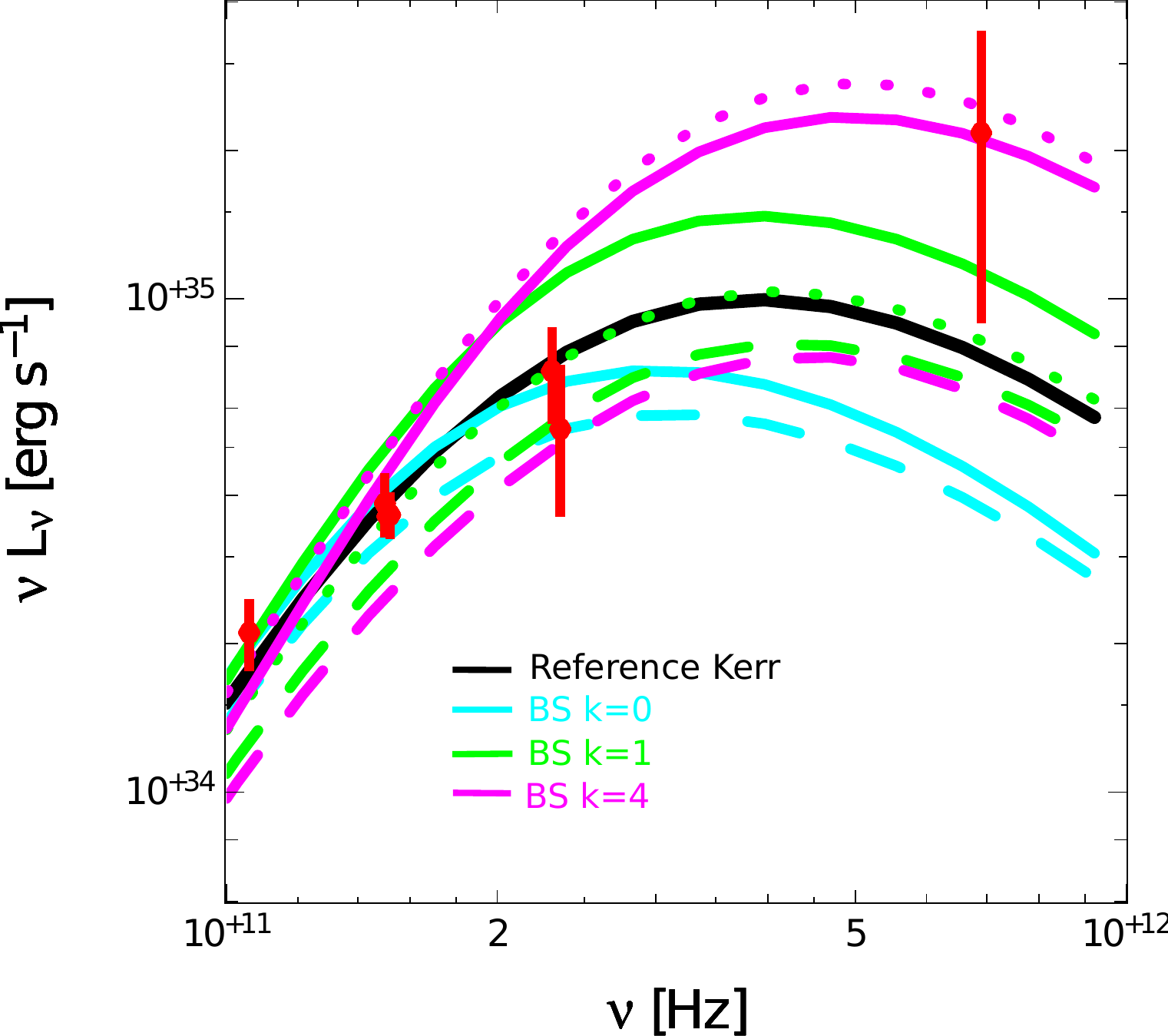}
	\caption{\textit{Comparing millimeter spectra}. The Kerr reference spectrum is in solid black.
	Boson stars (BS) spectra are in cyan for $k=0$, green for $k=1$ and magenta for $k=4$.
	Dotted lines are for $\omega = 0.7\,m/\hbar$, dashed lines for $\omega = 0.8\,m/\hbar$
	and solid lines for $\omega = 0.9\,m/\hbar$ ($\omega = 0.83\,m/\hbar$ for the $k=0$ case).
	We note the extreme similarity between the Kerr reference spectrum (solid black) and the
	$k=1$, $\omega = 0.7\,m/\hbar$ boson-star spectrum (dotted green), corresponding to the very similar strong-field
	images shown in Figure~\ref{fig:kerrref}, upper left panel, and Figure~\ref{fig:images}, middle right panel.}
	\label{fig:spec}
\end{figure}

Figure~\ref{fig:spec} shows the corresponding millimeter spectra for all boson-star spacetimes
as well as for the Kerr reference case.
It shows that different setups lead to different spectra. 
%However, given that the accretion tori themselves
%are somewhat different from spacetime to spacetime (due to the different $r_{\mathrm{in}}$), this difference should not be interpreted as only
%due to the differing geometry. 
However, it is not likely that spectra can provide a way to differentiate 
alternative compact objects given how degenerate the different parameters are. Taking different
values of the astrophysical parameters like the central density and temperature will lead to very different
spectra while the angular size of the "shadow" (be it the usual Kerr shadow or the faint central region 
in highly relativistic boson-star spacetimes) will not differ as it is due to lensing effects which are independent
of astrophysics. It is to be noticed still that the $(k=1,\omega=0.7\,m/\hbar)$ spectrum (dotted green) is
extremely similar to the Kerr reference spectrum (solid black): both the image and the spectra are thus extremely
similar to the Kerr case for this spacetime.
%We thus provide Figure~\ref{fig:spec} for completeness but do not believe it to be
%an important tool for differentiating compact objects.

\subsubsection{Photon orbit, bow-shape structure and spacetime stability}
\label{sec:stability}

The ($k=1$, $\omega=0.7$) spacetime we highlighted in the previous Section as able to generate
a Kerr-similar strong-field image may suffer from two stability issues. 

First, this solution is located
at $\omega<\omega_{\mathrm{max}}(k=1)$ as already written in the Introduction.
It is thus secularly unstable.

Second,~\citet{cardoso14} advocates the fact that all spacetimes with a stable photon orbit
and no event horizon are unstable. The ($k=1$, $\omega=0.7$) spacetime indeed has a stable
photon orbit. However, we believe that the statement of~\citet{cardoso14} is not sufficient
to be able to conclude with full confidence: a stability study of rotating boson-star spacetimes is
thus very much needed.

\begin{figure}
	\centering
	\includegraphics[height=7cm,width=8cm]{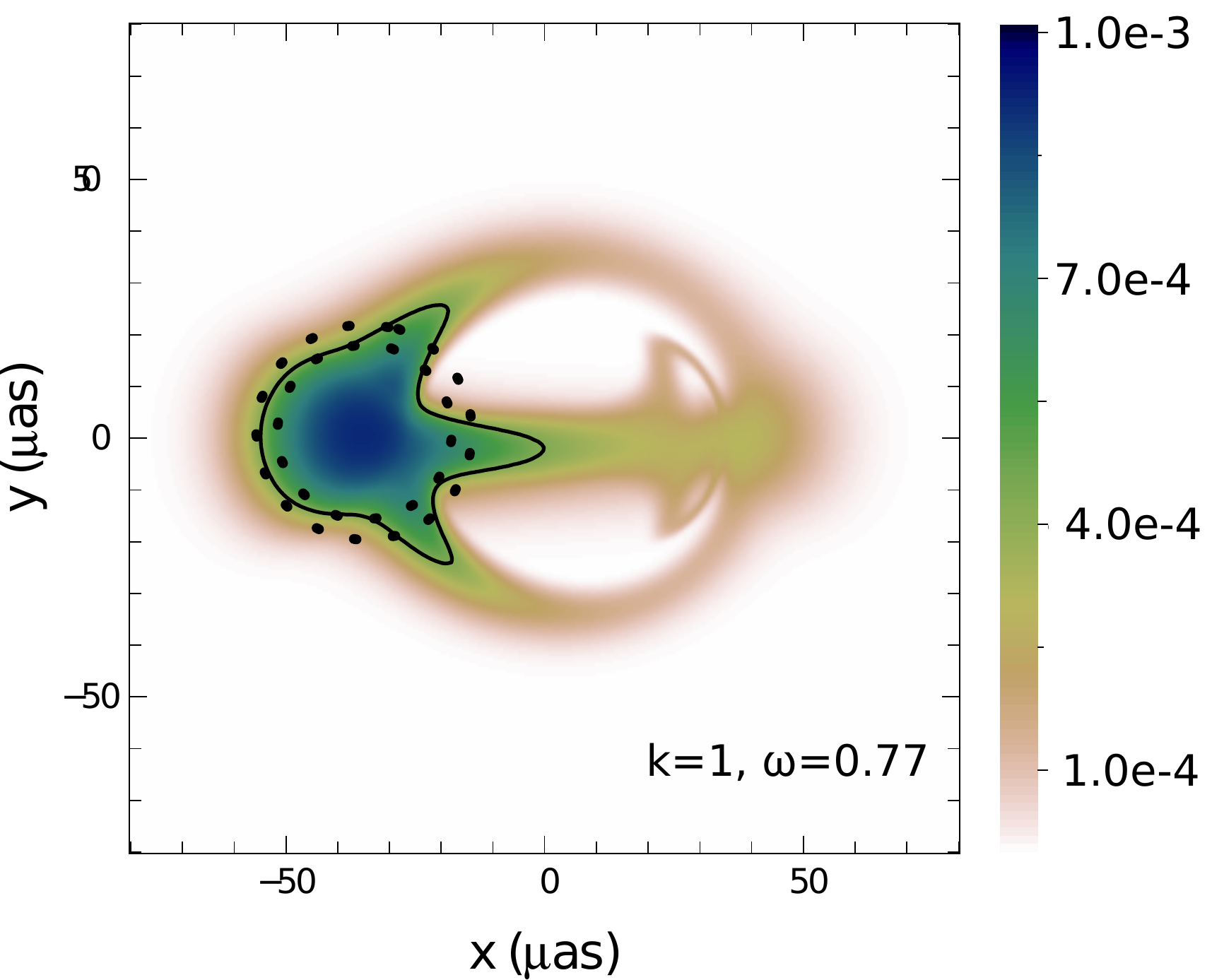}
	\caption{\textit{Bow-shape structure for a spacetime with no stability issue}. Image at $1.3$~mm
	for a $k=1$, $\omega=0.77$ boson star. This spacetime is most probably stable 
	as it is on the stable branch of the $M(\omega)$ curve, has no photon orbit and no ergoregion.
	It still displays a bow-shape structure, although it is smaller than in Fig.~\ref{fig:images}, middle right panel.}
	\label{fig:stability}
\end{figure}

Even if the $(k=1,\omega=0.7)$ spacetime may not be astrophysically relevant,
we consider that the fact that a spacetime with no event horizon can mimic a
Kerr strong-field image is sufficiently interesting to be highlighted. However, in order
to determine what the strong-field image will look like for a stable spacetime,
we have computed
one more image for $k=1$ boson stars, considering a frequency of $\omega=0.77$
(corresponding to the maximum of the $M(\omega)$ curve) which is
secularly stable. 
Moreover, the ($k=1$, $\omega=0.77$) spacetime has no photon orbit and
no ergoregion. There is thus to our knowledge no obvious reason to doubt its stability. 
The spin parameter of this configuration being $a=0.8$, it is also compatible with
a Kerr spacetime.
Figure~\ref{fig:stability}
shows a strong-field $1.3$~mm image for this spacetime. It still displays the bow-shape structure
typical of extremely strong lensing effect. This bow-shape structure, although smaller than in the
($k=1$, $\omega=0.7$) spacetime, is still similar to a portion of a Kerr photon ring. In particular, it appears
on the Doppler deboosted part of the image, which is the primary target for detecting photon
rings as highlighted by~\citet{psaltis14}.

Figure~\ref{fig:stability} thus shows that strong-field images with a clear decrease of intensity
in the central parts (a "shadow") and strong gradients of intensity (the bow-shape structure,
similar to a partly obscured photon ring) are not sufficient to tell an event horizon.

\section{Conclusion}
\label{sec:conclu}

We have performed ray-tracing computations of accretion tori surrounding Kerr black holes
and different kinds of boson stars in order to produce $1.3$~mm images and spectra of the accretion
flow surrounding Sgr~A* in the perspective
of future high-quality observations at this wavelength by the EHT. Our goal is to determine
how strong-field images differ from the well-known Kerr case when considering boson
stars, i.e. compact objects with no event horizon and no hard surface.

The main result of our research is Figure~\ref{fig:images} and particularly its central
right panel showing the image of an accretion torus around a $(k=1, \omega=0.7 \,m/\hbar)$
boson star which is extremely similar to a Kerr strong-field image. In particular, the image
shows a faint central region the angular size of which is very similar to that of the
Kerr shadow for the same spin and orientation. This finding questions the assumption
of~\citet{falcke00} and many other authors that detecting a shadow (i.e. a faint central region
separated by a strong intensity gradient from the exterior region) is a proof of the existence
of an event horizon.
Moreover, a bow-shape structure, due to
very strong light bending close to the center of the scalar field distribution, is visible in highly
relativistic boson-star spacetimes and is very similar to the Kerr photon ring.

Quite a few caveats should be noticed in order to interpret this result.
\begin{itemize}
\item A first obvious remark is that our model is stationary and made of a compact distribution
of normal matter which is not extending down to $r=0$ (which would be possible at least
in theory for
a boson-star spacetime given that there is no event horizon nor any singularity at $r=0$).
In case a long-lived accretion flow extending down to $r=0$ and emitting sufficiently
would be viable, it would not exhibit the same shadow-like central region. It is very difficult
to predict what such a flow would look like and we are now developing general relativistic magnetohydrodynamics numerical simulations
of such accretion structures in order to investigate this option.
\item We are considering in this paper mini boson stars (with no self-interactions among bosons), 
meaning that we have to assume the
existence of extremely light ($\approx 10^{-16}$~eV) spin-$0$ bosons in order to model Sgr~A*.
We plan to develop similar simulations as presented in this paper for self-interacting boson
stars that would allow modeling supermassive compact objects with a much higher boson mass.
We also note that~\citet{horvat13} studied boson stars non-minimally coupled to
gravity. This is another direction of generalization for the present work.
\item Our model assumes the stability of an accretion flow made of normal matter and surrounding a boson
star (for the typical parameters given in Table~\ref{tab:BS}). We are not aware
of any work studying in detail the evolution of baryonic matter around rotating boson 
stars, and in particular the possibility to form a black hole by accreting matter to $r=0$.
This is a very interesting area of research that we plan to investigate.
\item Finally, we have been assuming that normal matter does not interact with bosons
except through gravitational interaction.
%\textbf{[Is it useful to comment on this? Are there any alternatives studied?]}
\end{itemize}

However, despite all these limiting remarks, we believe that our result highlights
the extreme difficulty of interpreting strong-field images. In particular it shows
the importance, for the future interpretation of EHT data, 
of studying the observable predictions of well-established alternative compact objects,
in parallel to developing parameterized non-Kerr spacetimes. As highlighted by~\citet{cunha15}
it would be interesting to check whether these parameterized spacetimes can produce
such structures as the bow-shape feature exhibited in Figure~\ref{fig:images}.

As a final remark, we would like to stress that the aim of this article is \textit{not}
 to support the case for a boson star at the center of the Galaxy, or as an alternative to
 black hole candidates in general. Our aim is to investigate the simplest possible testbed
 of event-horizon-less spacetime. We believe that this simplest testbed is the boson-star model.
 As a consequence, boson stars are useful tools to investigate the power of experiments
 aiming at demonstrating the existence of black holes. Such experiments should first demonstrate
 their ability to tell a black hole from a boson star. This article shows that 
 experiments based on the investigation of shadows of compact objects may not be valid tests of the existence
 of black holes because it is not clear that they are able to unambiguously differentiate a black hole
 from a boson star. It is possible, although not clear at the moment, that gravitational-wave tests
 could be a clean way to differentiate a black hole from a boson star~\citep{ryan97,kesden05,palenzuela08}.

% ACKNOWLEDGMENTS %

\ack
FHV is grateful to Enrico Barausse for pointing out the work of~\citet{cardoso14} and to Carlos Herdeiro,
Helgi Runarsson
and Thomas Sotiriou for stimulating discussions at a mini-workshop on scalar fields
and gravitation in Meudon. 
FHV acknowledges financial support from the National Science Centre (NCN), Poland, grant 2013/09/B/ST9/00060
and was partially supported by the National Science Centre (NCN), Poland, DEC-2013/08/M/ST9/00664, 
within the framework of the HECOLS International Associated Laboratory.
Part of this work was supported by the French PNHE (Programme National Hautes \'Energies). 
ZM acknowledges financial support from the UnivEarthS Labex program at Sorbonne Paris Cit\'e (ANR-10-LABX- 0023 and ANR-11-IDEX-0005-02).
EG acknowledges support from the grant ANR-12-BS01-012-01 \emph{Analyse Asymptotique en Relativit\'e G\'en\'erale}.

%  APPENDIX %

%\appendix

%\section{Geodesic equation in Chern-Simons gravity} 

\bibliographystyle{aa}
\bibliography{ImagingBS}

\end{document}